\documentclass [twoside,12pt] {article}
\usepackage{setspace,graphicx,natbib,bm,fancyhdr}
\usepackage{amsmath,amsthm,amssymb,latexsym}
\usepackage{color}
\usepackage{graphicx}

\usepackage{authblk}

\newcommand{\rf}{\vskip .1in\par\sloppy\hangindent=1pc\hangafter=1
                 \noindent}
\newcommand{\chp}[1]{\mbox{$\stackrel{\wedge}{#1}$}}
\newcommand{\ichp}[1]{\mbox{$\stackrel{\vee}{#1}$}}

\newcommand{\vct}[1]{\mbox{$\stackrel{\rightharpoonup}{#1}$}}

\newcommand{\slas}[1]{\mbox{${{#1} \!\!\! /}$}}

\begin{document}
\title{\bf Second quantized quantum field theory based on invariance properties of locally conformally flat space-times} 
\author{John Mashford \\
School of Mathematics and Statistics \\
University of Melbourne, Parkville, Vic. 3010, Australia \\
E-mail: mashford@unimelb.edu.au}
\date{\today}
\maketitle

\begin{abstract}

Well defined quantum field theory (QFT) for the electroweak force including quantum electrodynamics (QED) and the weak force is obtained by considering  natural unitary representations of a group $K\subset U(2,2)$, where $K$ is locally isomorphic to $SL(2,{\bf C})\times U(1)$,  on a state space of Schwartz spinors, a Fock space ${\mathcal F}$ of multiparticle states and a space ${\mathcal H}$ of fermionic multiparticle states which forms a Grassmann algebra. These algebras are defined constructively and emerge from the requirement of covariance associated with the geometry of space-time. (Here $K$ is the structure group of a certain principal bundle associated with a given M\"{o}bius structure modeling space-time.) Scattering processes are associated with intertwining operators between various algebras,  which are encoded in an associated bundle of kernel algebras. Supersymmetry emerges naturally from the algebraic structure of the theory. Kernels can be generated using $K$ covariant matrix valued measures given a suitable definition of covariance. It is shown how Feynman propagators, fermion loops and the electron self energy can be given well defined interpretations as measures covariant in this sense. An example of the methods described in the paper is given in which the first order Feynman amplitude of electro-electron scattering ($ee\rightarrow ee$) is derived from a simple order (2,2) kernel. A second example is given explaining muon decay which is a manifestation of the weak force.

\end{abstract}

\section{Introduction}

Since the earliest development of quantum field theory (QFT), and to the dismay of its inventors such as Dirac and Feynman, QFT has been subject to problems associated with infinities or divergences in its defining equations. It has been found possible through the processes known as regularization and renormalization, to manipulate these infinities, {\em post factum} and produce answers in close agreement with experimental observation. However it has not been found possible to modify the defining equations or principles to give finite well defined governing equations which produce the same outputs or predictions as the result of regularizing and  renormalizing the divergent, not well defined, equations.

We propose that the fundamental reason for this state of affairs is that the Feynman propagators for the scalar particle, photon, fermion, W boson etc. are being treated as complex, complex $4\times4$ matrix or tensor valued functions defined on Minkowski space ${\bf R}^4$ while they should, however, be viewed as appropriate valued measures on ${\bf R}^4$. We propose that the basic objects of (2$^{\mbox{nd}}$ quantized) QFT should be measures or operators defined on function spaces such as the space of Schwartz spinors which we will define below. 

The question then arises as to which of such measures or operators have physical significance and our answer is that this comes about through the geometry and then subsequently the topology of space-time. In a previous paper (Mashford, 2017) it was shown how modeling space-time as a locally Minkowskian causal structure leads to the consideration of M\"{o}bius structures (which are closely related to locally conformally flat (pseudo-)Riemannian manifolds) and associated bundles. In particular there is associated with any M\"{o}bius structure a principal bundle $Q$ with structure group a group $K$. Classical quantum field theory (relativistic quantum mechanics)  as defined by Maxwell's equations and the Dirac equation were derived by considering actions of $K$ on ${\bf R}^4$ and ${\bf C}^4$ respectively. 

We will define a number of natural function algebras on which $K$ acts and consider bundles associated to $Q$ through these actions. It will be shown that the requirement that components of objects at any given point in space-time transform covariantly (in an appropriate sense) leads one to seek $K$ intertwining operators (operators which commute with the action of $K$).

 One can then consider bundles of algebras of such operators and seek distinguished fields by consideration of an analogue of the de Rham cohomology and its relation to space-time topology. 

\subsection{Related work}

Our work essentially provides an axiomatization of QFT using few axioms (e.g. that the space-time ``causal structure" is locally Minkowskian). Some of the earliest work on the axiomatization of QFT was done by Wightman in the 1950s (Streater and Wightman, 1989). One of the basic principles of their axiomatization is that the space of states in QFT is a Hilbert space on which the Poincar\'{e} group acts unitarily. In our work the space of states is not a Hilbert space in the strict mathematical sense because the Hermitian form  on it is not positive definite. Non- positive definite ``Hilbert spaces" have been used for a long time in QFT (e.g. the Gupta-Bleuler method), and it is generally realized that the metric on the space of states may not be positive definite. 

Furthermore in our work, the group which acts unitarily on the space of states is the group $K$ and $K$ is not isomorphic to the Poincar\'{e} group. 
The fact that the symmetry group of our theory is the group $K$ rather than the Poincar\'{e} group also implies that the Coleman Mandula theorem precluding the intermixing of space-time and internal symmetries may not apply.

The standard $SU(3)\times SU(2)\times U(1)$ model (SM) of strong, electromagnetic and weak interactions has properties associated with being conformally invariant since all the associated coupling constants are dimensionless, and gauge bosons and fermions get masses dynamically by spontaneous symmetry breaking. 
The SM exhibits ``near conformal invariance" (Meissner and Nicolai, 2007) with the lack of invariance being related to divergent terms at very high orders of precision. If the divergences could be ``cured" then 4D conformal invariance may manifest.

Models of 2D conformal field theory (CFT) (Gaberdiel, 2000) are associated with universality classes of the critical behaviour of 2D systems in statistical mechanics. In 2D CFT the method of conformal bootstrap derives all properties of the theory by requiring the consistency of a small set  of assumptions (Belavin and Tarnopolsky, 2010).

String Theory is an interesting development over the last few decades in which novel models are postulated for the underlying geometry of space-time and structures which it carries as is studied extensively in work of Green {\em et al.} (2012) and many others. There are 5 versions of string theory and it is thought that the string theories are 5 different perturbative expansions of a theory known as M-theory (Lambert, 2012; Becker, 2007). M-theory is a supersymmetric theory as is the theory that we will describe. However, unlike string theory, the outputs of our theory are Feynman amplitudes for the electroweak force and a clear path to the strong force seems apparent. 

Supersymmetry (SUSY) has been considered in QFT for a considerable length of time because of its possible benefit for eliminating divergences, its properties with respect to grand unified theories (GUTS) (since at 1 loop order the running inverse gauge couplings meet in MSSM (minimal supersymmetric standard model) but not in the SM (provided the superpartner masses are in a suitable range)), and for the possibility that it might provide a solution to the SM ``fine tuning problem".

\subsection{Summary of paper}

In the present paper a  bundle, whose typical fiber ${\mathcal H}_1$ is an infinite dimensional topological complex vector space with Hermitian form, associated to $Q$ by a unitary representation of the structure group $K$ of $Q$ is constructed. More generally, an algebra bundle with typical fiber isomorphic to the Fock space ${\mathcal F}=\bigoplus_{n=0}^{\infty}{\mathcal F}_n$, where ${\mathcal F}_n=\bigotimes_{i=1}^n{\mathcal H_1}$, for $n=1, 2, \ldots$ and ${\mathcal F_0}={\bf C}$ (the complex numbers), of multiparticle states is constructed. As is common in field theory we define the space of states (fermionic states) to be ${\mathcal H}=\bigoplus_{n=0}^{\infty}{\mathcal H}_n$ where ${\mathcal H}_n=\bigwedge_{i=1}^n{\mathcal H_1}$ for $n=1,2,\ldots$ and ${\mathcal H}_0={\bf C}$. 

Morphisms for ${\mathcal F}$ and ${\mathcal H}$ are intertwining operators and such intertwining operators can be constructed using $K$ covariant matrix valued measures. This form of covariance which is associated with the dual action of $K$ on ${\bf R}^4$ and ${\bf C}^4$ is shown to be satisfied by the scalar field, photon, electron and W boson propagators. Scattering processes with $k$ $|$in$>$ and $l$ $|$out$>$ particles are associated with linear maps which intertwine with the action of $K$ (morphisms) $\Xi:{\mathcal H}_k\rightarrow{\mathcal D}_l$ (where ${\mathcal D}_l$ is a certain space of smooth functions) which can be generated by order $(k,l)$ intertwining kernels.

It is shown how one can derive the Feynman amplitude for M\o ller electron-electron scattering ($ee\rightarrow ee$) from a simple order $(2,2)$ intertwining kernel. We also show that the Feynman amplitude for muon decay can be derived from another order $(2,2)$ intertwining kernel. We will propose that all the scattering processes for the electroweak force are associated with intertwining kernels.

It is also proposed that an analogue of the de Rham cohomology applied to the algebra of $K$ intertwining kernels would provide a description of ``actually occurring" scattering processes in terms of the topology of space-time.

\section{Second quantization}

\subsection{The space ${\mathcal H}_1$ of Schwartz spinors}

In (Mashford, 2017) we defined the principal bundle $Q$ associated with any given M\"{o}bius structure and investigated the physics arising from considering the fundamental representation of its structure group $K$ on ${\bf C}^4$ and the representation of $K$ on ${\bf R}^4$.

We now carry out what may be described as ``$2^{\mbox{nd}}$ quantization" by considering the action of $K$ on certain function spaces. In particular consider a state space ${\mathcal H}_1$ defined to be the set of functions $u\in C^{\infty}({\bf R}^4,{\bf C}^4)$ which, along with their derivatives have rapid decrease. Thus ${\mathcal H}_1={\mathcal S}({\bf R}^4,{\bf C}^4)$ is the appropriate analogue of the Schwartz class (Friedlander, 1982).

Specifically, we define seminorms $q_{k,l}:C^{\infty}({\bf R}^4,{\bf C}^4)\rightarrow[0,\infty]$ by
\begin{eqnarray}
q_{k,l}(u) & = & \mbox{sup}\{(1+|p|^2)^{\frac{k}{2}}|(D^{\alpha}u)(p)|:p\in{\bf R}^4,\alpha:\{0,1,2,3\}\rightarrow\{0,1,2,\ldots\} \nonumber \\
    &  &  \mbox{ is a multi-index }\mbox{with } |\alpha|\leq l\}, \nonumber
\end{eqnarray}
for $k,l\in\{0,1,2,\ldots\}$, in which $|\mbox{ . }|$ denotes the usual Euclidean norm on both ${\bf R}^4$ and ${\bf C}^4$ and also the usual multi-index norm, and then we define
\begin{equation}
{\mathcal H}_1=\{u\in C^{\infty}({\bf R}^4,{\bf C}^4):q_{k,l}(u)<\infty,\forall k,l\in\{0,1,2,\ldots\}\}.
\end{equation}
${\mathcal H}_1$ has the structure of a topological vector space (c.f. Friedlander, 1982). We will call elements of ${\mathcal H}_1$ Schwartz spinors. Schwartz spinors describe wave packets in Minkowski space. (Note that we do not restrict the spinor functions $u\in{\mathcal H}_1$ to be Dirac eigenstates, i.e. we do not assume that the test functions $u$ neccessarily satisfy the Dirac equation.)

In general, suppose that we have a group $G_1$ which acts on spaces $Z_1$ and $Z_2$. Let $Z_2^{Z_1}$ denote the set of all maps $f:Z_1\rightarrow Z_2$. Then $G_1$ acts on $Z_2^{Z_1}$ in a natural way according to
\begin{equation} \label{eq:natural_action}
(gf)(z)=gf(g^{-1}z), \forall g\in G_1, z\in Z_1.
\end{equation}
If we have a subset $T\subset Z_2^{Z_1}$ e.g. a collection of linear and/or continuous functions (where $Z_1$ and $Z_2$ have appropriate structures) such that $G_1T\subset T$, i.e. $gf\in T, \forall g\in G_1,f\in T$ then $T$ inherits the action defined above for $Z_2^{Z_1}$. 

We recall, from (Mashford, 2017), that the group $K\subset U(2,2)$ is defined by
\begin{equation}
K=\left\{\left(\begin{array}{cc}
a&0\\
0&a^{\dagger-1}
\end{array}\right):a\in GL(2,{\bf C}),|\mbox{det}(a)|=1\right\}.
\end{equation}
$K$ acts on both ${\bf R}^4$ and ${\bf C}^4$ in natural ways. Specifically 
\begin{eqnarray}
(\kappa,p) & \mapsto & (\kappa p=\Lambda(\kappa)p)\in{\bf R}^4\mbox{ for }\kappa\in K,p\in{\bf R}^4, \\
(\kappa,v) & \mapsto & \kappa v\in{\bf C}^4\mbox{ for }\kappa\in K,v\in{\bf C}^4,
\end{eqnarray}
where $\Lambda(\kappa)$ is the Lorentz transformation corresponding to $\kappa$ (Mashford, 2017). Therefore $K$ acts on ${\mathcal H}_1$ in a natural way according to
\begin{equation} \label{action}
(\kappa u)(p)=\kappa u(\kappa^{-1}p)=\kappa u(\Lambda(\kappa)^{-1}p), \forall \kappa\in K, p\in {\bf R}^4.
\end{equation}

The action of $K$ on ${\mathcal H}_1$ preserves the topology and vector space structure of ${\mathcal H}_1$. Shortly we will give ${\mathcal H}_1$ a Hermitian form (in fact a collection of Hermitian forms) and the Hermitian form is also preserved by the action. Therefore the action defines a unitary representation of $K$ on ${\mathcal H}_1$.

\subsection{Fock space ${\mathcal F}$, the space ${\mathcal H}$ of physical fermionic multiparticle states and the algebras ${\mathcal P}$, ${\mathcal Q}$, ${\mathcal C}$ and ${\mathcal D}$}

A single particle state at any given point in space-time may be described by a Schwartz spinor $u\in{\mathcal H}_1$. 

 A Schwartz spinor is a function from ${\bf R}^4$ to ${\bf C}^4$. It may be thought of as a map $u:(p,\alpha)\mapsto u(p,\alpha)\in{\bf C}$ where $p\in{\bf R}^4$ is the momentum variable (where the spinor is being viewed in momentum space) and $\alpha\in\{0,1,2,3\}$ is the polarization variable. Equivalently, it may be thought of as a vector valued function $u^{\alpha}:{\bf R}^4\rightarrow{\bf C}$, for $\alpha=0,1,2,3$.

Generalizing from this, an $n$ particle state may be thought of as a map $u:(p_1,\alpha_1,\ldots,p_n,\alpha_n)\mapsto u(p_1,\alpha_1,\ldots,p_n,\alpha_n)\in{\bf C}$ where $p_i\in{\bf R}^4, \alpha_i\in\{0,1,2,3\}$ for $i=1,\ldots,n$. Equivalently, an $n$ particle state may be thought of as a tensor valued function $u^{\alpha_1\ldots\alpha_n}:({\bf R}^{4})^n\rightarrow{\bf C}$ where the indices $\alpha_i$ all take values in $\{0,1,2,3\}$ for $i=1,\ldots,n$.  Denote by 
\begin{equation}
{\mathcal F}_n=\bigotimes_{i=1}^n{\mathcal H}_1={\mathcal H}_1\otimes\ldots\otimes{\mathcal H}_1, \mbox{ ($n$ times)}, 
\end{equation}
the space of such tensor valued functions which are Schwartz in all arguments, ${\mathcal F}_0={\bf C}$ and
\begin{equation}
{\mathcal F}=\bigoplus_{n=0}^{\infty}{\mathcal F}_n,
\end{equation}
the general Fock space of all fermion multiparticle states.

More generally define 
\begin{equation}
{\mathcal C}_n=\bigotimes_{i=1}^n{\mathcal C}_1={\mathcal C}_1\otimes\ldots\otimes{\mathcal C}_1, \mbox{ ($n$ times)}, 
\end{equation}
where ${\mathcal C}_1=C^{\infty}({\bf R}^4,{\bf C}^4)$,
the space of all smooth tensor valued functions $u^{\alpha_1\ldots\alpha_n}\in C^{\infty}(({\bf R}^4)^n,{\bf C})$,  ${\mathcal C}_0={\bf C}$ and
\begin{equation}
{\mathcal C}=\bigoplus_{n=0}^{\infty}{\mathcal C}_n.
\end{equation}

Define, for $r,s=1,2,\ldots,$ a function $f:{\bf R}^r\rightarrow{\bf C}^s$ to be polynomially bounded if there exists a polynomial function $P:{\bf R}^r\rightarrow{\bf C}^s$ such that 
\begin{equation}
|f(x)|\leq|P(x)|, \forall x\in{\bf R}^r.
\end{equation}
Denote the space of polynomially bounded functions  $u\in C^{\infty}({\bf R}^4,{\bf C}^4)$ by ${\mathcal P}_{1}$ and let
\begin{equation}
{\mathcal P}=\bigoplus_{n=0}^{\infty}{\mathcal P}_n,
\end{equation}
where ${\mathcal P}_0={\bf C}$ and, for $n=1,2,\ldots$
\begin{equation}
{\mathcal P}_n=\bigotimes_{i=1}^{n}{\mathcal P}_1 = {\mathcal P}_1\otimes\ldots\otimes{\mathcal P}_1, \mbox{ ($n$ times)},
\end{equation}  
is the space of all smooth tensor valued functions $u^{\alpha_1\ldots\alpha_n}\in C^{\infty}(({\bf R}^{4})^n,{\bf C})$ which are polynomially bounded in all arguments. Then clearly ${\mathcal F}_n\subset{\mathcal P}_n\subset{\mathcal C}_n, \forall n=0,1,2,\ldots$ and ${\mathcal F}\subset{\mathcal P}\subset{\mathcal C}$.

$K$ acts on ${\mathcal H}_1$ according to 
\begin{equation}
(\kappa u)^{\alpha}(p)={\kappa^{\alpha}}_{\beta}u^{\beta}(\kappa^{-1}p)={\kappa^{\alpha}}_{\beta}u^{\beta}(\Lambda(\kappa)^{-1}p), \forall\kappa\in K, p\in{\bf R}^4,
\end{equation}
(summing over repeated indices). The generalization to Fock space is clear. $K$ acts on ${\mathcal F}_n$ according to
\begin{eqnarray} \label{eq:action}
(\kappa u)^{\alpha_1\ldots\alpha_n}(p_1,\ldots,p_n) & = & {\kappa^{\alpha_1}}_{\beta_1}\ldots{\kappa^{\alpha_n}}_{\beta_n}u^{\beta_1\ldots\beta_n}(\kappa^{-1}p_1,\ldots,\kappa^{-1}p_n), \forall\kappa\in K,  \nonumber \\
   &  & p_1,\ldots,p_n\in{\bf R}^4.
\end{eqnarray}
These actions extend to actions of $K$ on ${\mathcal P}_n$ and ${\mathcal C}_n$ for $n=1,2,\ldots$ (using the same formula Eq.~\ref{eq:action}).

We also define ${\mathcal H}$ to be $\bigoplus_{n=0}^{\infty}{\mathcal H}_n$ where ${\mathcal H}_0={\bf C}$ and for $n=1,2,\ldots,{\mathcal H}_n=\bigwedge_{i=1}^n{\mathcal H_1}={\mathcal H}_1\wedge\ldots\wedge{\mathcal H}_1$ ($n$ times) and we call  ${\mathcal H}$ the space of physical fermionic states. Here $\bigwedge_{i=1}^n{\mathcal H}_1$ is defined as follows.
\begin{eqnarray}
{\bigwedge}_{i=1}^n{\mathcal H}_1 & = & \{u\in{\mathcal F}_n:u^{\pi(\alpha)}(\pi(p))=\epsilon_{\pi} u^{\alpha}(p), \\
    &  & \forall\alpha\in\{0, 1 ,2, 3\}^n, p\in({\bf R}^4)^n, \pi\in S_n\}, \nonumber
\end{eqnarray}
where $S_n$ denotes the set of all permutations of $\{1,\ldots,n\}$, $\epsilon_{\pi}$ denotes the signature of the permutation $\pi\in S_n$ and, for $\pi\in S_{n}, \pi(\alpha)=(\alpha_{\pi(1)},\ldots,\alpha_{\pi(n)})$ for any index vector $\alpha=(\alpha_1,\ldots,\alpha_n)\in\{0,1,2,3\}^n$ and $\pi(p)=(p_{\pi(1)},\ldots,p_{\pi(n)})$ for any $p\in({\bf R}^4)^n$. 
(As in the case of the usual definition of the exterior algebra, ${\mathcal H}$ can be given an equivalent definition as a quotient of ${\mathcal F}$ by a certain two sided ideal.)

Additionally, we define the space ${\mathcal Q}$ to be  $\bigoplus_{n=0}^{\infty}{\mathcal Q}_n$ where ${\mathcal Q}_0={\bf C}$ and, for $n=1,2,\ldots,{\mathcal Q}_n=\bigwedge_{i=1}^n{\mathcal P_1}={\mathcal P}_1\wedge\ldots\wedge{\mathcal P}_1$ ($n$ times). Here $\bigwedge_{i=1}^n{\mathcal P}_1$ is defined as follows.
\begin{eqnarray}
{\bigwedge}_{i=1}^n{\mathcal P}_1 & = & \{u\in{\mathcal P}_n:u^{\pi(\alpha)}(\pi(p))=\epsilon_{\pi} u^{\alpha}(p), \\
    &  & \forall\alpha\in\{0, 1 ,2, 3\}^n, p\in({\bf R}^4)^n,\pi\in S_n\}, \nonumber
\end{eqnarray}
and, similarly, we define the space ${\mathcal D}=\bigoplus_{n=0}^{\infty}{\mathcal D}_n$ where ${\mathcal D}_0={\bf C}$ and, for $n=1,2,\ldots$, ${\mathcal D}_n=\bigwedge_{i=1}^n{\mathcal C}_1$.

The Fock space ${\mathcal F}$ and the spaces ${\mathcal P}$ and ${\mathcal C}$ form graded algebras under tensor multiplication $(u,v)\mapsto u\otimes v$ if we define for $u=((p_1,\alpha_1,\ldots,p_n,\alpha_n)\mapsto u^{\alpha_1\ldots\alpha_n}(p_1,\ldots,p_n))$ and $v=((q_1,\beta_1,\ldots,q_m,\beta_m)\mapsto v^{\beta_1\ldots\beta_m}(q_1,\ldots,q_m))$ the object $u\otimes v$ defined by 
\begin{equation}
(u\otimes v)^{\alpha_1\ldots\alpha_{n}\beta_1\ldots\beta_m}(p_1,\ldots,p_{n},q_1,\ldots,q_m)=u^{\alpha_1\ldots\alpha_n}(p_1,\ldots,p_n)v^{\beta_{1}\ldots\beta_{m}}(q_{1},\ldots,q_{m}).
\end{equation}
By a tedious but straightforward calculation it can be shown that the spaces ${\mathcal H}$, ${\mathcal Q}$ and ${\mathcal D}$ form graded algebras under the exterior (Grassmann) multiplication $\wedge$ defined by
\begin{equation}
(u\wedge v)^{\alpha}(p)=\frac{1}{(n+m)!}\sum_{\pi\in S_{n+m}}\epsilon_{\pi}(u\otimes v)^{\pi({\alpha})}(\pi(p)),
\end{equation}
for $u\in{\mathcal H}_n, v\in{\mathcal H}_ m$, $u\in{\mathcal Q}_n, v\in{\mathcal Q}_ m$ or $u\in{\mathcal D}_n, v\in{\mathcal D}_ m$.  Thus, in other words, the operations $\wedge:{\mathcal H}\times{\mathcal H}\rightarrow{\mathcal H}$, $\wedge:{\mathcal Q}\times{\mathcal Q}\rightarrow{\mathcal Q}$ and $\wedge:{\mathcal D}\times{\mathcal D}\rightarrow{\mathcal D}$ are distributive as well as being linear. (The aforementioned tedious calculation shows that $\wedge$ is distributive given triples of elements from ${\mathcal H}_k, {\mathcal H}_l$ and ${\mathcal H}_m$ and we extend it distributively to all of ${\mathcal H}$, similarly for ${\mathcal Q}$ and ${\mathcal D}$.)

Clearly ${\mathcal H}_n\subset{\mathcal Q}_n\subset{\mathcal D}_n, \forall n=0,1,2,\ldots$ and ${\mathcal H}\subset{\mathcal Q}\subset{\mathcal D}$.
It is easy to show that
\begin{equation}
u\wedge v=-v\wedge u, \forall u,v\in{\mathcal D}_1,
\end{equation}
and so
\begin{equation}
u\wedge u=0, \forall u\in{\mathcal D}_1.
\end{equation}

Given a space $Y$ on which $K$ acts by automorphisms let $F(Y)$ denote the bundle associated to $Q$ through the action of $K$ on $Y$.

It is straightforward to show that $K$ acts on ${\mathcal F},{\mathcal P},{\mathcal C},{\mathcal H},{\mathcal Q}$ and ${\mathcal D}$  by graded algebra automorphisms.
Therefore $F({\mathcal F}),F({\mathcal P}),F({\mathcal C}),F({\mathcal H}),F({\mathcal Q})$ and $F({\mathcal D})$ have structures of bundles of graded algebras. 

\subsection{Some general fiber bundle theory}

If $Y$ is a space on which $K$ acts and $x\in X$ (space-time) let $F(Y)_x$ denote the fiber $\pi^{-1}(x)$ of the fiber bundle $F(Y)$ where $\pi:F(Y)\rightarrow X$ is the canonical projection.
An element $v\in F(Y)_x$ may be viewed as a map $v:I_x\rightarrow Y$ which satisfies
\begin{equation}
v_i=\kappa_{ij}(x)v_j, \forall i,j\in I_x,
\end{equation}
where $\{\kappa_{ij}\}_{i,j\in I}$ are the transition functions for $Q$, $I_x=\{i\in I:x\in U_i\}$ and ${\mathcal A}=\{(\phi_i,U_i,V_i)\}_{i\in I}$ is the atlas for $X$.

If $Y$ and $Z$ are spaces on which $K$ acts let $F(Y,Z,x )$ denote the set of maps $f:F(Y)_x\rightarrow F(Z)_x$. If $Y$ and $Z$ have the structures of vector spaces and the actions of $K$ on them are linear then $F(Y)$ and $F(Z)$ are vector bundles and we let 
\begin{equation}
L(Y,Z,x)=\{f\in F(Y,Z,x):f \mbox{ is linear}\}.
\end{equation}

We are interested in characterizing the set of maps $F(Y,Z,x)$ or, at least, finding a natural class of objects which induce elements of $F(Y,Z,x)$. Suppose that $\Xi:Y\rightarrow Z$ is an intertwining map (for the given actions of $K$). i.e. 
\begin{equation}
\Xi(\kappa y)=\kappa\Xi(y), \forall\kappa\in K, y\in Y.
\end{equation}
Define $f=f_{\Xi}:F(Y)_x\rightarrow Z^{I_x}$ by
\begin{equation} \label{eq:invariant_function1}
f(v)_i=f_{\Xi}(v)_i=\Xi(v_i).
\end{equation}
Then, for all $v\in F(Y)_x$
\begin{equation}
f(v)_i=\Xi(v_i)=\Xi(\kappa_{ij}v_j)=\kappa_{ij}\Xi(v_j)=\kappa_{ij}f(v)_j,\forall i,j\in I_x,
\end{equation}
and so $f(v)\in F(Z)_x$. Thus $f\in F(Y,Z,x)$. Therefore we have determined a way that elements of $F(Y,Z,x)$ can be constructed from intertwining maps from $Y$ to $Z$.

Let $\mbox{Int}(Y,Z)$ denote the set of intertwining maps from $Y$ to $Z$. Then, given an element $\Xi\in\mbox{Int}(Y,Z)$, Eq.~\eqref{eq:invariant_function1} defines an element $f_{\Xi}\in F(Y,Z,x)$. In fact, the map $\Xi\mapsto f_{\Xi}$ is an imbedding of $\mbox{Int}(Y,Z)$ in $F(Y,Z,x).$ 

Suppose that $Y$ and $Z$ have structures of vector spaces and the actions of $K$ on them are linear. Let $L(Y,Z)$ denote the space of linear mappings from $Y$ to $Z$. Then if $\Xi\in L(Y,Z)\cap\mbox{Int}(Y,Z)$ is a linear intertwining operator then $f_{\Xi}\in L(Y,Z,x)$. Furthermore $ L(Y,Z)\cap\mbox{Int}(Y,Z)$ is a vector space and $\Xi\mapsto f_{\Xi}$ is an imbedding of the vector space   $L(Y,Z)\cap\mbox{Int}(Y,Z)$ in the vector space $L(Y,Z,x)$, for any $x\in X$.

\subsection{Particle scattering and the S matrix}

Now $F({\mathcal F}_k)_x$ is the Fock space of $k$-multiparticle states at the point $x\in X$. It would be reasonable to represent a multiparticle scattering process at $x$ with $k$ $|$in$>$ and $l$ $|$out$>$ particles by a linear map from $F({\mathcal F}_k)_x$ to $F({\mathcal F}_l)_x$. We find it convenient to give a more general definition of a scattering process at $x$ with $k$ $|$in$>$ and $l$ $|$out$>$ particles as being represented by a linear map from $F({\mathcal F}_k)_x$ to $F({\mathcal P}_l)_x$ or, even more generally, as a linear map from $F({\mathcal F}_k)_x$ to $F({\mathcal C}_l)_x$. This generalization is partially justified by the fact that all the $K$ invariant Hermitian forms $<\mbox{ },\mbox{ }>=(\mbox{ },\mbox{ }):{\mathcal F}_n\times{\mathcal F}_n\rightarrow{\bf C}$ that we will define in Section~\ref{section:Hermitian_forms} extend to sesquilinear forms $<\mbox{ },\mbox{ }>:{\mathcal F}_n\times{\mathcal P}_n\rightarrow{\bf C}$. Thus if we have a $(k,k)$ particle scattering process defined by an intertwining operator $\Xi:{\mathcal F}_k\rightarrow{\mathcal P}_k$ then we can still compute
\begin{equation}
<\mbox{in}|\mbox{out}>=<\mbox{in}|\Xi|\mbox{in}>.
\end{equation} 
 
Thus we propose that particle scattering processes at a point $x\in X$ are associated with  linear mappings from the fiber $F({\mathcal F}_{k})_{x}$ in the vector bundle $F({\mathcal F}_k)$ to the fiber $F({\mathcal P}_l)_x$ of the vector bundle $F({\mathcal P}_l)$ (or, more generally, to the fiber $F({\mathcal C}_l)_x$ of the vector bundle $F({\mathcal C}_l)$).

Let, for $k,l\in\{1,2,\ldots\}$,
\begin{eqnarray}
{\mathcal I}_{kl}& = & L({\mathcal F}_k,{\mathcal C}_l)\cap{\mathcal I}({\mathcal F}_k,{\mathcal C}_l) \\
    & = & \{\Xi:{\mathcal F}_k\rightarrow{\mathcal C}_l\mbox{ such that $\Xi$ is a linear intertwining operator}\}. \nonumber
\end{eqnarray}
We have shown above that every element of ${\mathcal I}_{kl}$ generates a scattering process at any given point in space-time and so we may call  operators $\Xi\in{\mathcal I}_{kl}$ S matrices or S operators.

A bundle homomorphism from $F({\mathcal F}_k)$ to $F({\mathcal C}_l)$ is a continuous map $f:F({\mathcal F}_k)\rightarrow F({\mathcal C}_l)$ whch takes each fiber $F({\mathcal F}_k)_x$ linearly into the fiber $F({\mathcal C}_l)_x$ for all $x\in X$. It follows from the above that any continuous map $\theta:X\rightarrow{\mathcal I}_{kl}$ induces a bundle homomorphism, which describes $(k,l)$ scattering processes for each point in space-time.

\section{Superspace and superfields  \label{section:superfields}}

The  concept of supersymmetry (Buchbinder and Kuzenko, 1998; Arnowitt {\em et al.}, 1984) plays an important part in modern approaches to understanding physics at a fundamental level. We will show that supersymmetry is present given certain algebraic preconditions in a theory. In usual treatments in the physics literature supersymmetry is imposed {\em ad hoc}. In our work, to be described below, we show that the required preconditions are present in our theory and we therefore derive, or deduce, the presence of supersymmetry. In other words we do not arbitrarily impose supersymmetry properties (such as ${\bf Z}_2$-gradings) to the algebraic structures of our theory, we derive them.

Let, for $V$ and $W$ complex vector spaces, $\mbox{Mult}_{n}(V,W)$ denote the space of $n$-multilinear maps from $V^n=V \times ... \times V$ ($n$ times) to $W$ and let $\mbox{Alt}_{n}(V,W)$ denote the space of alternating multilinear maps in $\mbox{Mult}_{n}(V,W)$ for $n=1,2,\ldots$ and Alt$_0(V,W)=$ Mult$_0(V,W)={\bf C}$.

Suppose that we have an action $v\mapsto\kappa v$ of $K$ on a complex vector space $V$. Suppose also that $A$ is an algebra over ${\bf C}$ on which $K$ acts. 
$K$ acts on Alt$_n(V,A)$ according to
\begin{equation}
<\kappa\Xi;v_{1}, \ldots , v_{n}> =  \kappa<\Xi;\kappa^{-1} v_{1}, \ldots , \kappa^{-1} v_{n}>.
\end{equation}
Here we are using the notation $<\mbox{ };\mbox{ }>$ to denote the evaluation map. (The action is the natural action on Alt$_n(V,A)$ induced, according to Eq.~\ref{eq:natural_action} by the action $(\kappa;v_1,\ldots,v_n)\mapsto(\kappa v_1,\ldots,\kappa v_n)$ of $K$ on $V^n$ and the action of $K$ on $A$.) 

Now there is a natural product $(\Xi_1,\Xi_2)\mapsto\Xi_1\wedge\Xi_2=\Xi_1\Xi_2$ of $p$-forms in Alt$_{p}(V,A)$ by $q$-forms in Alt$_{q}(V,A)$ to produce $n$ forms in Alt$_{n}(V,A)$,  where $p,q\in\{1,2,\ldots\}$ and $n = p+q$, defined as follows.
\newtheorem{definition}{Definition}
\begin{definition}
If $\Xi_1 \in \mbox{Alt}_p(V,A)$ and
 $\Xi_2 \in \mbox{Alt}_q(V,A)$ then the product $\Xi_1\Xi_2=\Xi_1\wedge\Xi_2$ of $\Xi_1$ and $\Xi_2$ is given by
\begin{equation}
<\Xi_{1}\Xi_{2}; v_{1}, \ldots , v_{n}> = \frac{1}{n!}\sum_{\sigma\in S_{n}}
\epsilon_{\sigma}<\Xi_{1};v_{\sigma(1)}, \ldots , v_{\sigma(p)}>
<\Xi_{2}; v_{\sigma(p+1)}, \ldots , v_{\sigma(n)}>,
\end{equation}
for all $v_1,\ldots,v_n\in V.$
\end{definition}
This definition is a straightforward generalization to algebra valued forms of the product for scalar valued forms (Greub {\em et al.}, 1972, Volume I, p. 5). To show that this product is well defined we prove the following.
\newtheorem{proposition}{Proposition}
\begin{proposition} \label{alternating_forms}
If $\Xi_1 \in \mbox{Alt}_p(V,A)$ and
 $\Xi_2 \in \mbox{Alt}_q(V,A)$ then the product $\Xi_1\Xi_2$ of $\Xi_1$ and $\Xi_2$ is an element of Alt$_n(V,A)$.
\end{proposition}
{\bf Proof}
It is clear that $\Xi_1\Xi_2$ is a well defined $A$ valued multilinear form. We need to show that it is an alternating form. To this end we compute as follows.
\begin{eqnarray}
  & & <\Xi_1\Xi_2;v_{\rho(1)},\ldots,v_{\rho(n)}> \nonumber \\
  & = & 
\frac{1}{n!}\sum_{\sigma\in S_n}\epsilon_{\sigma} <\Xi_1;v_{\rho(\sigma(1))},\ldots,v_{\rho(\sigma(p))}>
<\Xi_2;v_{\rho(\sigma(p+1))},\ldots,v_{\rho(\sigma(n))}> \nonumber \\
  & = &
\frac{1}{n!}\sum_{\sigma\in S_n}\epsilon_{(\rho^{-1}\sigma)} <\Xi_1;v_{\sigma(1)},\ldots,v_{\sigma(p)}>
<\Xi_2;v_{\sigma(p+1)},\ldots,v_{\sigma(n)}> \nonumber \\
  & = &
\epsilon_{\rho}\frac{1}{n!}\sum_{\sigma\in S_n}\epsilon_{\sigma} <\Xi_1;v_{\sigma(1)},\ldots,v_{\sigma(p)}>
<\Xi_2;v_{\sigma(p+1)},\ldots,v_{\sigma(n)}> \nonumber \\
  & = &
\epsilon_{\rho}<\Xi_1\Xi_2;v_1,\ldots,v_n>, \nonumber
\end{eqnarray}
for all $v_1,\ldots,v_n\in V,\rho\in S_n$. $\Box$

Let 
\begin{equation}
\mbox{Alt}(V,A) = \bigoplus_{n=0}^{\infty} \mbox{Alt}_{n}(V,A).
\end{equation}
Now
\begin{equation} \label{eq:graded}
\mbox{Alt}_n(V,A) \mbox{Alt}_m(V,A) \subset \mbox{Alt}_{n+m}(V,A), \forall n,m=0,1,2,\ldots.
\end{equation}
Therefore Alt$(V,A)$ is a graded algebra over ${\bf C}$ (Greub {\em et al.}, 1972, p. 3).

Now $K$ acts by linear isomorphisms on Alt$(V,A)$. However, in general $K$ does not act by algebra isomorphisms on Alt$(V,A)$. Let, for $n=1,2,\ldots,{\mathcal J}_n(V,A)$ denote the set of $K$ invariant elements of Alt$_n(V,A)$ i.e.
\begin{equation}
 {\mathcal J}_n(V,A)=\{\Xi\in\mbox{Alt}_n(V,A):\kappa\Xi=\Xi,\forall\kappa\in K\},
\end{equation}
${\mathcal J}_0(V,A)={\bf C}$ and ${\mathcal J}(V,A)$ denote the graded algebra 
\begin{equation}
{\mathcal J}(V,A)=\bigcup_{n=0}^{\infty}{\mathcal J}_n(V,A).
\end{equation}
Then (trivially) $K$ acts by algebra automorphisms on ${\mathcal J}(V,A)$.

Let, for each $n=1,2,\ldots, S^{(n)}$ denote the vector bundle associated to $Q$ by the action of $K$ on ${\mathcal J}_n(V,A)$ and $S$ be the graded algebra bundle defined by
\begin{equation}
S=\bigcup_{n=0}^{\infty}S^{(n)}.
\end{equation}
Let $\pi_{n}:S^{(n)}\rightarrow X$ and $\pi : S\rightarrow X$ be the canonical projections. For each $x\in X$, $\pi^{-1}(x)$ has the structure of a graded algebra. 

We can also make $\pi^{-1}(x)$ into a Lie algebra by defining the following natural bracket operation. For a $p-$form 
$\Xi_{1}\in \mbox{Alt}_{p}(V,A)$ and a $q-$form
$\Xi_{2}\in \mbox{Alt}_{q}(V,A)$ define $[\Xi_{1},\Xi_{2}]$ to be the $n = p+q$ form in Alt$_{n}(V,A)$ defined by
\begin{equation}
<[\Xi_{1},\Xi_{2}];v_{1}, \ldots , v_{n}> = \frac{1}{n!}\sum_{\sigma\in S_{n}}\epsilon_{\sigma}[<\Xi_{1};v_{\sigma(1)}, \ldots , v_{\sigma(p)}>,<\Xi_{2};v_{\sigma(p+1)}, \ldots ,v_{\sigma(n)}>],
\end{equation}
where $[\mbox{ },\mbox{ }]:A\times A\rightarrow A$ is the usual Lie bracket operation for $A$. 
After we have checked the appropriate analogue of Proposition~\ref{alternating_forms} we know that $[\mbox{ , }]$ defines a well defined operation on Alt$_n(V,A)\times$Alt$_m(V,A)$. Now
\begin{equation}
[\mbox{Alt}(V,A)_n,\mbox{Alt}(V,A)_m]\subset \mbox{Alt}(V,A)_{n+m},\forall n,m=0,1,2,\dots.
\end{equation}
Therefore Alt$(V,A)$ has the structure of a graded Lie algebra.
\nopagebreak
Now in order to determine the relationship between the (Grassmann) multiplication $\wedge$ and the Lie algebra operation in Alt$(V,A)$ we make the following computation.
\[ \begin{array}{l}
<[\Xi_{1},\Xi_{2}]; v_{1}, \ldots , v_{n}> \\
\mbox{ } \\
= \frac{1}{n!}\sum_{\sigma\in S_{n}}\epsilon_{\sigma}[<\Xi_{1};v_{\sigma(1)}, \ldots , v_{\sigma(p)}>,<\Xi_{2}; v_{\sigma(p+1)}, \ldots , v_{\sigma(n)} > ] \\
\mbox{ } \\
=\frac{1}{n!}\sum_{\sigma\in S_{n}}\epsilon_{\sigma}(
<\Xi_{1}; v_{\sigma(1)}, \ldots , v_{\sigma(p)}><\Xi_{2}; v_{\sigma(p+1)}, \ldots , v_{\sigma(n)}> - \\
<\Xi_{2}; v_{\sigma(p+1)}, \ldots , v_{\sigma(n)}><\Xi_{1}; v_{\sigma(1)}, \ldots , v_{\sigma(p)}>) \\
\mbox{ } \\
=\frac{1}{n!}\sum_{\sigma\in S_{n}}\epsilon_{\sigma}
<\Xi_{1}; v_{\sigma(1)}, \ldots , v_{\sigma(p)}><\Xi_{2}; v_{\sigma(p+1)}, \ldots , v_{\sigma(n)}> -  \\
\frac{1}{n!}\sum_{\sigma\in S_{n}}\epsilon_{\sigma}
<\Xi_{2}; v_{\sigma(p+1)}, \ldots , v_{\sigma(n)}><\Xi_{1}; v_{\sigma(1)}, \ldots , v_{\sigma(p)}> \\
\mbox{ } \\
=\frac{1}{n!}\sum_{\sigma\in S_{n}}\epsilon_{\sigma}
<\Xi_{1}; v_{\sigma(1)}, \ldots , v_{\sigma(p)}><\Xi_{2}; v_{\sigma(p+1)}, \ldots , v_{\sigma(n)}> -  \\
\frac{1}{n!}\sum_{\sigma\in S_{n}}\epsilon_{\sigma}(-1)^{pq}
<\Xi_{2}; v_{\sigma(1)}, \ldots , v_{\sigma(q)}><\Xi_{1}; v_{\sigma(q+1)}, \ldots , v_{\sigma(n)}>  \\
\mbox{ } \\
= <\Xi_{1}\Xi_{2} - (-1)^{pq}\Xi_{2}\Xi_{1}; v_{1}, \ldots , v_{n}> . 
\end{array} \]
Therefore
\begin{equation}
[\Xi_{1},\Xi_{2}] = \Xi_{1}\Xi_{2} - (-1)^{pq}\Xi_{2}\Xi_{1},
\end{equation}
which is the fundamental defining characteristic of supersymmetry theory in terms of commutation and anticommutation from the point of view of the universal construction of superalgebras (Buchbinder and Kuzenko, 1998, p.~123). Thus, for each $x\in X, \pi^{-1}(x)$ has the structure of a supersymmetry algebra and $S$ can be considered as a bundle of supersymmetry algebras. We will call the bundle $S$ superspace.

We will shortly show why it is natural to consider an S operator (S matrix) in QFT as an intertwining operator $\Xi:{\mathcal H}_k\rightarrow{\mathcal D}_l$ in which case there is induced an alternating multilinear map $T_{\Xi}:{{\mathcal H}_1}^k\rightarrow {\mathcal D}$ according to
\begin{equation}
T_{\Xi}(u_1,\ldots,u_k)=\Xi(u_1\wedge\ldots\wedge u_k).
\end{equation}
We will show that $T_{\Xi}\in{\mathcal J}_k({\mathcal H}_1,{\mathcal D})$ and so the machinery of supersymmetry developed above applies.

Define a {\em superfield} of degree $n$ for $n = 1, 2, \ldots$ to be a collection $\Psi = \{ \Psi_{i} \}_{i\in I}$ of alternating $A$ valued form fields $\Psi_{i}:V_{i}\rightarrow\mbox{Alt}_{n}(V,A)$ (continuous) which transform according to,
\begin{eqnarray}
<\Psi_{i}(\xi);v_{1}, \ldots , v_{n}> & = & (\kappa_{ij}\circ\phi_i^{-1})(\xi)<(\Psi_{j}\circ\phi_{j}\circ\phi_{i}^{-1})(\xi); \nonumber \\
    &  &  (\kappa_{ij}\circ\phi_{i}^{-1})(\xi)^{-1}v_{1},\ldots,(\kappa_{ij}\circ\phi_{i}^{-1})(\xi)^{-1}v_{n}>, \nonumber
\end{eqnarray}
$\forall\xi\in V_i=\phi_i(U_i), i\in I, v_1,\ldots,v_n\in V$ (where the atlas ${\mathcal A}$ for $X$ is given by ${\mathcal A}=\{(U_i,\phi_i)\}_{i\in I}$). 
Define a superfield of degree $0$ to be a collection
$\Psi = \{ \Psi_{i} \}_{i\in I}$ of ${\bf C}$ valued fields $\Psi_{i}:V_{i}\rightarrow {\bf C}$ which transform according to,
\begin{eqnarray}
\label{superfield_0}
\Psi_{i}(\xi) & = & (\Psi_{j}\circ\phi_{j}\circ\phi_{i}^{-1})(\xi).
\end{eqnarray}
Let ${\cal S}^{(n)}$ denote the space of superfields of degree $n$ and let, 
\begin{equation}
{\cal S} = \bigoplus_{n=0}^{\infty}{\cal S}^{(n)}.
\end{equation}
Superfields $\Psi_{n}\in{\cal S}^{(n)}$ of degree $n$ are precisely the sections of the bundle $S^{(n)}$. Superfields $\Psi\in{\cal S}$ are precisely the sections of the superspace bundle $S$.

\section{Feynman propagators as tempered measures}

In this section we give well defined definitions of Feynman propagators of QFT in terms of tempered measures and distributions. Propagators are viewed in a new way as being Lie algebra valued tempered distributions. They are constructed from Fourier transforms of  $u(2,2)$ valued tempered measures.

\subsection{The Feynman scalar field propagator \label{Feynman}}

 Consider the Feynman scalar field propagator. It is written as (Itzykson and Zuber, 1980, p. 35)
\begin{equation}
\label{meson1}
\triangle_{F}(x) = -(2\pi)^{-4}\int \frac{e^{-ip.x}}{p^{2}-m^{2}+i\epsilon}\, dp.
\end{equation}
This is to be understood with respect to the $i-$epsilon procedure described in Mandl and Shaw (Mandl and Shaw, 1984, p. 57), and the dot product $p.x$ is given by 
\[ p.x = \eta_{\alpha\beta} p^{\alpha} x^{\beta}, \]
where $\eta = \mbox{diag}(1,-1,-1,-1)$. Therefore $\triangle_{F}(x)$ is written as
\begin{equation}
\label{meson2}
\triangle_{F}(x) = -(2\pi)^{-4}\int_{{\bf R}^{3}} \int_{C_{F}} \frac{e^{-ip.x}}{p^{2}-m^{2}}\, dp^{0} d{\vct p},
\end{equation}
where $C_{F}$ is the standard Feynman propagator contour. Thus $\triangle_{F}(x)$ is written as
\begin{equation}
\label{meson3}
\triangle_{F}(x) = -(2\pi)^{-4}\int_{{\bf R}^{3}} I({\vct p},x)\, d{\vct p},
\end{equation}
where
\begin{equation}
\label{meson4}
I({\vct p},x) = \int_{C_{F}} \frac{e^{-ip.x}}{(p^{0})^{2}-\omega_{m}({\vct p})^{2}}\, dp^{0},
\end{equation}
and
\begin{equation}
\label{meson5}
\omega_{m}({\vct p}) = (|{\vct p}|^{2}+m^{2})^{\frac{1}{2}}, m\geq 0.
\end{equation}
The contour integral \ref{meson4} exists for $\omega_{m}({\vct p}) \neq 0$ and is given by 
\begin{equation}
\label{meson6}
I({\vct p},x) = -\frac{\pi i}{\omega_{m}({\vct p})}\left\{ \begin{array}{cc}
e^{-i(\omega_{m}({\vct p})x^{0}-{\vct p}.{\vct x})} & \mbox{ if } x^{0} > 0, \\
e^{-i(-\omega_{m}({\vct p})x^{0}-{\vct p}.{\vct x})} & \mbox{ if } x^{0} < 0.
\end{array} \right.
\end{equation}
To prove this 
consider the contour $C_1(R)$ given by 
\[ C_1(R) = \{ Re^{it}:0\leq t \leq \pi\}. \]
We will show that
\[ I_1(R) = \int_{C_1(R)}\frac{e^{-ip.x}}{(p^0)^2-\omega_m(\vct{p})^2}\,dp^0\rightarrow 0 \mbox{ as } R\rightarrow \infty, \]
as long as $x^0 < 0$. To this effect we note that
\begin{eqnarray}
|I_1(R)| &=& \left|\int_{t=0}^{\pi}\frac{e^{-iRe^{it}x^0+i\vct{p}.\vct{x}}}{(Re^{it})^2-\omega_m(\vct{p})^2}iRe^{it}\,dt\right| \nonumber \\
&\leq& \int_{t=0}^{\pi}\left|\frac{e^{R\sin tx^0}}{(Re^{it})^2-\omega_m(\vct{p})^2}\right|R\,dt \nonumber \\
&\leq&\int_{t=0}^{\pi}\frac{1}{\left|R^2-\omega_m(\vct{p})^2\right|}R\, dt \nonumber \\
&=& \frac{\pi R}{\left|R^2-\omega_m(\vct{p})^2\right|} \nonumber \\
&& \rightarrow 0 \mbox{ as }R\rightarrow\infty, \mbox{ if } x^0<0. \nonumber
\end{eqnarray}
Therefore, for $x^0<0$,
\[ I(\vct{p},x) = 2\pi i\mbox{res}(p^0 \mapsto \frac{e^{-ip.x}}{(p^0)^2-\omega_m(\vct{p})^2},-\omega_m(p)).\]
Now
\[ \frac{e^{-ip.x}}{(p^0)^2-\omega_m(\vct{p})^2}=\frac{e^{-ip.x}}{(p^0-\omega_m(\vct{p}))(p^0+\omega_m(\vct{p}))}.\]
Thus
\[ I(\vct{p},x) = -\pi i\frac{e^{-i(-\omega_m(\vct{p})x^0-\vct{p}.\vct{x})}}{\omega_m(\vct{p})}. \]
Similarly, if $x^0>0$, then
\[ I(\vct{p},x) = -\pi i \frac{e^{-i(\omega_m(\vct{p})x^0-\vct{p}.\vct{x})}}{\omega_m(\vct{p})}. \] 
Hence 
\begin{equation}
\label{meson7}
\int_{{\bf  R}^{3}} |I({\vct p},x)|\, d{\vct p} = \pi\int_{{\bf R}^{3}}\frac{1}{\omega_{m}({\vct p})}\, d{\vct p} = \infty,
\end{equation}
and so the integral \ref{meson3} defining $\triangle_{F}(x)$ does not exist as a Lebesgue integral.

We would like to give a well defined interpretation of the propagator $\triangle_{F}$. To do this let the hyperboloids $H_{m}^{+}, H_{m}^{-} \subset {\bf R}^{4}$ for $m > 0$ and the cones $H_{0}^{+}$ and $H_{0}^{-}$ be defined by
\begin{equation}
\label{meson8}
H_{m}^{\pm} = \{ (\pm \omega_{m}({\vct p}),{\vct p}) : {\vct p}\in{\bf R}^{3} \}, m > 0,
\end{equation} 
\begin{equation}
H_{0}^{\pm} = \{ (\pm \omega_{0}({\vct p}),{\vct p}) : {\vct p}\in{\bf R}^{3}\setminus \{0\} \}. 
\end{equation}
$H_{m}^{\pm}$ for $m \ge 0$ are orbits of the action of the Lorentz group on Minkowski space (these orbits correspond to real mass orbits, there are also ``imaginary mass" hyperboloid orbits).
Let ${\mathcal B}({\bf R}^4)$ denote the Borel algebra for ${\bf R}^4$.
Define $\Omega_{m}^{\pm} : {\mathcal B}({\bf R}^{4}) \rightarrow [0,\infty]$ by
\begin{equation}
\label{meson9}
\Omega_{m}^{\pm}(\Gamma) = \int_{\pi(\Gamma \cap H_{m}^{\pm})} \frac{d{\vct p}}{\omega_{m}({\vct p})},
\end{equation}
where $\pi : {\bf R}^{4} \rightarrow {\bf R}^{3}$ is defined by
\begin{equation}
\label{meson10}
\pi(p^{0},{\vct p}) = {\vct p}.
\end{equation}
Then $\Omega_{m}^{\pm}$ are Lorentz invariant measures for Minkowski space supported on $H_{m}^{\pm}$ (Bogolubov {\em et al.}, 1975, p. 157). $\Omega_{m}^{\pm}$ is locally finite for $m\ge 0$.
Now, for any non-negative measurable function $\psi : {\bf R}^{4} \rightarrow [0,\infty]$,
\begin{equation}
\label{meson11}
\int_{{\bf R}^{4}} \psi(p)\, \Omega_{m}^{\pm}(dp) = \int_{{\bf R}^{3}} \psi(\pm\omega_{m}({\vct p}),{\vct p})\frac{d{\vct p}}{\omega_{m}({\vct p})}.
\end{equation}

Here, and for the rest of the section, the symbol $\psi$ stands for a test function in Minkowski space.

It follows from Equations~\ref{meson6} and \ref{meson11} that one may write
\begin{equation}
\label{meson12}
\triangle_{F}(x) = \left\{ \begin{array}{cc}
(2\pi)^{-4}\pi i\int e^{-ip.x}\, \Omega_{m}^{+}(dp), & \mbox{ if } x^{0} > 0, \\
(2\pi)^{-4}\pi i\int e^{-ip.x}\, \Omega_{m}^{-}(dp), & \mbox{ if } x^{0} < 0.
\end{array} \right.
\end{equation}
The Equations \ref{meson2}, \ref{meson3} and \ref{meson12} are all integral expressions equivalent to \ref{meson1} and none of them exist as Lebesgue integrals. However, formally, Equation \ref{meson12} can be written as
\begin{equation}
\label{meson13}
\triangle_{F}(x) = \left\{ \begin{array}{cc}
\pi i {\ichp {\Omega_{m}^{+}}}(-x) & \mbox{ if } x^{0} > 0, \\
\pi i {\ichp {\Omega_{m}^{-}}}(-x) & \mbox{ if } x^{0} < 0,
\end{array} \right.
\end{equation}
where $ ^\vee$ denotes the (distributional) inverse Fourier transform operator. Since $\Omega_m^{+}$ and $\Omega_m^{-}$ are tempered distributions their Fourier transforms exist and are tempered distributions.
Let ${\cal S}^{\pm}({\bf R}^{4})\subset {\cal S}({\bf R}^{4})$ be the space of test functions supported in $S^{\pm}$, where
\begin{equation}
S^{+} = \{x\in {\bf R}^4 : x^0 > 0\}, S^{-} = \{x \in {\bf R}^4 : x^0 < 0\}.
\end{equation}
Then 
\begin{equation}
<\triangle_F,\psi> =\pi i<{\ichp {\Omega_m^{\pm}}},\psi> =\pi i<\Omega_m^{\pm},{\ichp \psi}>,
\end{equation}
for $\psi\in{\cal S}^{\mp}({\bf R}^4)$,
where $<\omega,\psi>$ denotes the evaluation of a distribution $\omega$ on its test function argument $\psi$.
Therefore the momentum space scalar field propagator on $({\cal S}^{\mp}({\bf R}^4))^{\vee}$ is
\begin{equation}
{\chp \triangle}_F = \pi i\Omega_m^{\pm}.
\end{equation}
$({\cal S}^{+}({\bf R}^{4}))^{\vee}$ is the space of wave functions with only positive frequency components while 
$({\cal S}^{-}({\bf R}^{4}))^{\vee}$ is the space of wave functions with only negative frequency components.

This measure is a tempered measure, i.e. it is a tempered distribution as well as being a measure. (Trivially) it is $u(2,2)$ valued.

\subsection{The Feynman fermion field propagator}

We now turn from the Feynman scalar field propagator to the Feynman fermion propagator.
The Feynman fermion propagator is written as (Itzykson and Zuber, 1980, p. 150)
\begin{equation}
\label{fermion1}
S_{F}(x) = (2\pi)^{-4} \int \frac{{\slas p}+m}{p^{2}-m^{2}+i\epsilon} e^{-ip.x} dp,
\end{equation}
where ${\slas p}$ denotes the Feynman slash of $p$ defined by
\[ {\slas p} = p_{\mu}\gamma^{\mu}, \]
in which $\{\gamma^{\mu}\}_{\mu = 0}^3$ are the Dirac gamma matrices (which we take to be in the chiral representation in which case $i\gamma^{\mu}\in su(2,2),\forall\mu=0,1,2,3$, (Mashford, 2017)).
Arguing as with the scalar field propagator as in Equation~\ref{meson12} this can be written as
\begin{equation}
\label{fermion2}
S_{F}(x) = -\left\{ \begin{array}{cc}
(2\pi)^{-4}\pi i\int e^{-ip.x}\, ({\slas p}+m)\Omega_{m}^{+}(dp), & \mbox{ if } x^{0} > 0, \\
(2\pi)^{-4}\pi i\int e^{-ip.x}\, ({\slas p}+m)\Omega_{m}^{-}(dp), & \mbox{ if } x^{0} < 0.
\end{array} \right.
\end{equation}
Consider $\Omega_{f,m}^{\pm}$ written as
\begin{equation}
\Omega_{f,m}^{\pm}(\Gamma) = \int_{\Gamma} ({\slas p}+m)\Omega_{m}^{\pm}(dp).
\end{equation}
If $\Gamma$ is bounded then $\Omega_{f,m}^{\pm}(\Gamma)$ exists. $\Omega_{f,m}^{\pm}$ is a measure on ${\mathcal B}_0({\bf R}^4)=\{\Gamma\in{\mathcal B}({\bf R}^4):\Gamma\mbox{ is relatively compact}\}$ where ${\mathcal B}({\bf R}^4)$ denotes the Borel algebra of  ${\bf R}^4$. However $\Omega_{f,m}^{\pm}(\Gamma)$ may not exist for $\Gamma$ unbounded. $\Omega_{f,m}^{\pm}$ exists as a matrix valued tempered distribution (Choquet-Bruhat {\em et al.}, 1989, p. 476) according to 
\begin{equation}
\label{fermion3}
<\Omega_{f,m}^{\pm},\psi> = \int \psi(p)({\slas p}+m)\Omega_{m}^{\pm}(dp),
\end{equation}
for $\psi\in {\cal S}({\bf R}^{4})$ where ${\cal S}({\bf R}^{4})$ is the Schwartz space of test functions. Then $i\Omega_{f,m}^{\pm}$ is a $u(2,2)$ valued tempered distribution. Formally, Equation~\ref{fermion2} may be written as
\begin{equation}
\label{fermion4}
S_{F}(x) = -\left\{ \begin{array}{cc}
\pi i {\ichp {\Omega_{f,m}^{+}}}(-x) & \mbox{ if } x^{0} > 0, \\
\pi i {\ichp {\Omega_{f,m}^{-}}}(-x) & \mbox{ if } x^{0} < 0.
\end{array} \right.
\end{equation}
Let ${\cal S}^{\pm}({\bf R}^{4})\subset {\cal S}({\bf R}^{4})$ be the space of test functions supported in $S^{\pm}$. Then 
\begin{equation}
<S_{F},\psi> =-\pi i <{\ichp {\Omega_{f,m}^{\pm}}}, \psi>
     =  -\pi i < \Omega_{f,m}^{\pm}, {\ichp \psi}>,
\end{equation}
for $\psi\in {\cal S}^{\mp}({\bf R}^{4})$. Therefore the momentum space Feynman fermion propagator on \newline $({\cal S}^{\mp}({\bf R}^{4}))^{\vee}$ is
\begin{equation}
\label{fermion_propagator}
{\chp S_{F}} = -\pi i\Omega_{f,m}^{\pm}.
\end{equation}
Clearly this measure is a $u(2,2)$ valued tempered measure.
 
\subsection{The Feynman photon propagator}

The Feynman photon propagator is written as (Itzykson and Zuber, 1980, p. 133)
\begin{equation}
G_{F\mu\nu}(x) = -(2\pi)^{-4}\eta_{\mu\nu}\int \frac{e^{-ik.x}}{k^{2}+i\epsilon} \, dk,
\end{equation}
where, as for the scalar field propagator, the $i-\epsilon$ procedure is adopted (Mandl and Shaw, 1984, p. 57). 
Arguing as above, the Feynman photon propagator in momentum space is
\begin{equation}
{\chp G}_{F\mu\nu} =\eta_{\mu\nu}\pi i \Omega_0^{\pm},
\end{equation}
on $({\cal S}^{\mp}({\bf R}^4))^{\vee}$.

\subsection{The Feynman W boson propagator}

The Feynman W boson propagator is written as (Mandl and Shaw, 1991, p. 244)
\begin{equation}
D_F^{\alpha\beta}(x,m_W) = (2\pi)^{-4}\int \frac{-g^{\alpha\beta}+k^{\alpha}k^{\beta}/m_W^2}{k^2-m_W^2+i\epsilon}e^{-ik.x}dk,
\end{equation}
where, again, this is to be understood through the $i-$epsilon procedure in which the integral is a contour integral of the function
\[ h(x) = \frac{-g^{\alpha\beta}+k^{\alpha}k^{\beta}/m_W^2}{k^2-m_W^2}e^{-ik.x}, \]
over the contour $C_F$ defined previously for the scalar field propagator.
Arguing as above the Feynman W boson propagator in momentum space on ${\cal S}^{\mp}({\bf R}^4))^{\vee}$ is 
\begin{equation}
D_F^{\alpha\beta} = -\pi i(\Omega_{b,m_W}^{\pm})^{\alpha\beta},
\end{equation}
where $\Omega_{b,m_W}^{\pm}$ are the tensor valued tempered distributions defined by
\begin{equation}
<\Omega_{b,m_W}^{\pm},\psi>^{\alpha\beta} = \int \psi(k)(-g^{\alpha\beta}+k^{\alpha}k^{\beta}/m_W^2)\Omega_{m_W}^{\pm}(dk).
\end{equation}

\section{Covariance of propagator measures}

For $m\in {\bf R}\setminus\{0\}$ define
\begin{equation}
\Omega_m=\left\{\begin{array}{l}
\Omega_m^{+}\mbox{ if }m> 0 \\
\Omega_{-m}^{-}\mbox{ if }m<0
\end{array}\right.
\end{equation}
Then $\Omega_m$ is a Lorentz invariant measure. 

A matrix valued measure $\mu:{\mathcal B}_0({\bf R}^4)\rightarrow{\bf C}^{4\times 4}$ will be said to be $K$ covariant if
\begin{equation}
\mu(\kappa\Gamma)=\kappa\mu(\Gamma)\kappa^{-1}, \forall\kappa\in K, \Gamma\in{\mathcal B}_0({\bf R}^4).
\end{equation}
A matrix valued tempered distribution $\mu:{\mathcal S}({\bf R}^4,{\bf C})\rightarrow{\bf C}^{4\times 4}$ will be said to be $K$ covariant if
\begin{equation}
<\mu,\kappa\psi>=\kappa<\mu,\psi>\kappa^{-1}, \forall \kappa\in K, \psi\in{\mathcal S}({\bf R}^4,{\bf C}).
\end{equation}
The following theorem can be readily established.
\newtheorem{theorem}{Theorem}
\begin{theorem}
Let  $\mu$ be a matrix valued tempered measure. Then $\mu$ is $K$ covariant as a distribution  if and only if it is $K$ covariant as a measure.
\end{theorem}
Also it is straightforward to prove the following. 
\begin{theorem}
Suppose that $\mu:{\mathcal B}_0({\bf R}^4)\rightarrow{\bf C}^{4\times4}$ is a ${\bf C}^{4\times4}$ valued $K$ covariant tempered measure. Then
\begin{equation}
\mu*(\kappa u)=\kappa(\mu* u), \forall\kappa\in K,u\in{\mathcal H}_1,
\end{equation}
where $*$ denotes convolution, i.e.
\begin{equation}
(\mu* u)(p)=\int\mu(dq)u(p-q).
\end{equation}
\end{theorem}
Note that $\mu*u$ defined by the above equation exists and is in $C^{\infty}({\bf R}^4,{\bf C}^4)$. It follows that if $\mu$ is a $K$  covariant tempered measure then the operator $\Xi$ defined by
\begin{equation}
\Xi(u)=\mu* u, \mbox{ for } u\in{\mathcal H}_1,
\end{equation}
is an intertwining operator from ${\mathcal H}_1$ to ${\mathcal C}_1.$

Clearly $\Omega_m^{\pm}$ is a covariant measure for all $m\geq 0$ (because it is invariant as a scalar valued measure). Also we have the following.
\begin{theorem}
$i\Omega_{f,m}^{\pm}$ is a $K$ covariant $u(2,2)\subset{\bf C}^{4\times4}$ valued measure for all $m>0$.
\end{theorem}
{\bf Proof} 
\begin{eqnarray}
<i\Omega_{f,m}^{\pm},\kappa\psi> & = & i\int\psi(\kappa^{-1}p)({\slas p}+m)\Omega_m^{\pm}(dp) \nonumber \\
    & = & i\int\psi(p)(\Sigma(\kappa p)+m)\Omega_m^{\pm}(dp) \nonumber \\
    & = & i\int\psi(p)(\kappa{\slas p}\kappa^{-1}+m)\Omega_m^{\pm}(dp) \nonumber \\
    & = & \kappa i\int\psi(p)({\slas p}+m)\Omega_m^{\pm}(dp)\kappa^{-1} \nonumber \\
    & = & \kappa<i\Omega_{f,m}^{\pm},\psi>\kappa^{-1},  \nonumber
\end{eqnarray}
for all $\kappa\in K$ and measurable $\psi:{\bf R}^4\rightarrow{\bf C})$ which are $\Omega_{f,m}^{\pm}$ integrable (e.g. all $\psi\in{\mathcal S}({\bf R}^4,{\bf C})$ are so integrable). We have noted before that $i\Omega_{f,m}^{\pm}$ is $u(2,2)$ valued.
$\Box$

Define, because of its relation to the W boson propagator, the tensor valued tempered measure 
\begin{equation}K_{\mbox{weak}}^{\alpha\beta}(\Gamma)=\int_{\Gamma} p^{\alpha}p^{\beta}\,\Omega_m(dp).
\end{equation}
$K_{\mbox{weak}}$ is covariant under the natural action of the group $K$ on the space of measures on  ${\bf R}^4$ and the tensor space $T_{20}({\bf R}^4)$. 

It is straightforward to show that if $\mu:{\mathcal S}({\bf R}^4,{\bf C})\rightarrow{\bf C}^{4\times 4}$ is a $K$ covariant matrix valued tempered distribution then the Fourier transform ${\chp \mu}$ of $\mu$ (applied componentwise) is a $K$ covariant matrix valued tempered distribution.

Let $\sigma:{\mathcal B}({\bf R})\rightarrow{\bf C}$ be a Borel complex measure. Then we may define the measure $\mu=\mu_{\sigma}$ by
\begin{equation}
\mu(\Gamma)=\int_{m=-\infty}^\infty \Omega_m(\Gamma)\,\sigma(dm).
\end{equation}
It is straightforward to show that $\mu$ is a $K$ covariant Borel measure.
Similarly we may construct many $K$ covariant measures from the fermion propagator measure, the measure associated with the photon propagator and the weak force measure, using Borel measures on ${\bf R}$. 
The measure $\sigma$ in these constructions can be described as the mass spectrum of the constructed $K$ covariant measure.

\section{Unitary representations of $K$ on certain Hilbert spaces and spaces equipped with Hermitian forms \label{section:Hermitian_forms}}

Let $\sigma:{\mathcal B({\bf R})}\rightarrow [0,\infty]$ be a Borel measure.

Let 
\begin{eqnarray}
{\mathcal X}_{\sigma} & = & \{\psi:{\bf R}^4\rightarrow{\bf C}\mbox{ such that $\psi$ is Borel measurable and } \nonumber \\
    &  & \int_{m=-\infty}^{\infty}\int \psi(p)^{*}\psi(p)\,\Omega_m(dp)\,\sigma(dm)<\infty\}.
\end{eqnarray}
Define $(\mbox{ },\mbox{ }):{\mathcal X}_{\sigma}\times{\mathcal X}_{\sigma}\rightarrow{\bf C}$ by
\begin{equation}
(\psi^{(1)},\psi^{(2)})=\int_{m=-\infty}^{\infty}\int\psi^{(1)*}(p)\psi^{(2)}(p)\,\Omega_m(dp)\,\sigma(dm).
\end{equation}
Then $(\mbox{ },\mbox{ })$ is an inner product which makes ${\mathcal X}_{\sigma}$ into a Hilbert space on which $K$ acts by unitary transformations.

$\sigma$ will be said to be polynomially bounded if there exists a polynomial function $P:{\bf R}\rightarrow{\bf R}$ such that
\begin{equation}
\sigma(\Gamma)\leq\int_{\Gamma}|P(m)|\,dm, \forall\Gamma\in{\mathcal B}({\bf R}).
\end{equation}

Now, suppose that $\sigma$ is a polymomially bounded Borel measure. Then $K$ acts by unitary transformations on ${\mathcal H}_1={\mathcal S}({\bf R}^4,{\bf C}^4)$ when it is equipped with the Hermitian form
\begin{equation} \label{eq:inner_pdt}
(u,v)=\int_{m=-\infty}^{\infty}\int \overline{u}(p)v(p)\,\Omega_m(dp)\,\sigma(dm),
\end{equation}
where
\begin{equation}
\overline{u}(p)=\overline{u(p)}=u(p)^{\dagger}\gamma^{0}=u(p)^{\dagger}g,
\end{equation}
and $g$ is the Hermitian form for $U(2,2)$ which, by a fortuitous coincidence, coincides (in the chiral representation for the Dirac gamma matrices) with the gamma matrix $\gamma^0$ used to define conjugate spinors in QFT. However ${\mathcal H}_1$ with this Hermitian form does not form a Hilbert space in the strict (mathematical) sense because the Hermitian form is not positive definite. However it is well known in the physics community that the ``inner product" in the physical space of states may not be positive definite, however one nevertheless calls it a Hilbert space.

A particular Hermitian form for ${\mathcal H}_1$ arises because the Lebesgue measure is Lorentz invariant (since $|\mbox{det}(\Lambda)|=1, \forall \Lambda\in O(1,3)$), $K$ acts by unitary transformations on ${\mathcal H}_1$ when it is equipped with the Hermitian form
\begin{equation}
(u,v)=\int\overline{u}(p)v(p)\,dp,
\end{equation}
It can be shown that this is the Hermitian form generated by Eq.~\ref{eq:inner_pdt} when the spectral measure $\sigma$ is given by
\begin{equation}
\sigma(\Gamma)=\int_{\Gamma}|m|\,dm, \forall\Gamma\in{\mathcal B}({\bf R}).
\end{equation}
More generally we can define a $K$ invariant Hermitian form $(\mbox{ },\mbox{ }):{\mathcal F}_n\times{\mathcal F}_n\rightarrow{\bf C}$ for ${\mathcal F}_n$ by
\begin{eqnarray} 
(u,v) & = & \int u^{\alpha_1\ldots\alpha_n}(p_1,\ldots,p_n)^{*}(\gamma^0)_{\alpha_1\beta_1}\ldots(\gamma^0)_{\alpha_n\beta_n}v^{\beta_1\ldots\beta_n}(p_1,\ldots,p_n) \\
    &  & \,dp_1\ldots dp_n, \nonumber \\
    & = & \int u^{\alpha_1\ldots\alpha_n}(p_1,\ldots,p_n)^{*}v_{\alpha_1\ldots\alpha_n}(p_1,\ldots,p_n)\,dp_1\ldots dp_n. \nonumber
\end{eqnarray}
Thus evaluation of the Hermitian form involves complex conjugation of the first argument and ``lowering of the indices'' of the second argument by the metric tensor $g=\gamma^0$.

With further more generality we can define a $K$ invariant Hermitian form $(\mbox{ },\mbox{ }):{\mathcal F}_n\times{\mathcal F}_n\rightarrow{\bf C}$ by
\begin{eqnarray} 
(u,v) & = & \int_{m\in{\bf R}^n}\int u^{\alpha_1\ldots\alpha_n}(p_1,\ldots,p_n)^{*}v_{\alpha_1\ldots\alpha_n}(p_1,\ldots,p_n)\, \\
    &  & \Omega_{m_1}(dp_1)\ldots\Omega_{m_n}(dp_n)\,\sigma(dm), \nonumber
\end{eqnarray}
in which $\sigma:{\mathcal B}({\bf R}^n)\rightarrow[0,\infty]$ is a Borel polynomially bounded multiparticle mass spectrum.

All of the Hermitian forms on ${\mathcal F}_n$ that we have defined extend from $K$ invariant Hermitian forms on ${\mathcal F}_n\times{\mathcal F}_n$ to $K$ invariant sesquilinear forms on ${\mathcal F}_n\times{\mathcal P}_n$.

For $n\in\{1,2,\ldots\}$ define $F:{\mathcal F}_n\rightarrow{\mathcal F}_n$ (fermionization) by 
\begin{equation} \label{eq:fermionization}
(F(u))^{\alpha}(p)=\frac{1}{n!}\sum_{\pi\in S_n}\epsilon_{\pi}u^{\pi(\alpha)}(\pi(p)),
\end{equation}
for any index vector $\alpha\in\{0,1,2,3\}^{n}$ and $p\in({\bf R}^{4})^n$. 
Also define $B:{\mathcal F}_n\rightarrow{\mathcal F}_n$ (bosonization) by 
\begin{equation} \label{eq:bosonization}
(B(u))^{\alpha}(p)=\frac{1}{n!}\sum_{\pi\in S_n}u^{\pi(\alpha)}(\pi(p)).
\end{equation}
Then $F({\mathcal F}_n)={\mathcal H}_n,  F({\mathcal F}_n)\cap B({\mathcal F}_n)=\{0\}$.
Also
\begin{equation}
u\wedge v=F(u\otimes v), \forall u\in{\mathcal H}_k, v\in{\mathcal H}_l.
\end{equation}
$F$ and $B$ are idempotent and extend to idempotent operators $F:{\mathcal P}_n\rightarrow{\mathcal P}_n$, $B:{\mathcal P}_n\rightarrow{\mathcal P}_n$, $F:{\mathcal C}_n\rightarrow{\mathcal C}_n$ and $B:{\mathcal C}_n\rightarrow{\mathcal C}_n$ defined by Eqns.~\ref{eq:fermionization} and \ref{eq:bosonization}. $u\in{\mathcal C}_n$ will be said to be fermionic if $F(u)=u$ and bosonic if $B(u)=u$.

\section{$K$ intertwining operators from ${\mathcal F}_k$ to ${\mathcal P}_l$ defined by integral kernels}

We are interested in finding $K$ intertwining operators $\Xi:\mbox{$|$in$>$}\,\mapsto\mbox{$|$out$>$}$ mapping ${\mathcal F}_k$ to ${\mathcal P}_l$. Consider the class of operators defined by kernels of the form
\begin{eqnarray}
(\Xi(u))^{\alpha_1^{\prime}\ldots\alpha_l^{\prime}}(p_1^{\prime},\ldots,p_l^{\prime}) & = & \int {{\mathcal M}^{\alpha_1^{\prime}\ldots\alpha_l^{\prime}}}_{\alpha_1\ldots\alpha_k}(p_1^{\prime},\ldots,p_l^{\prime},p_1,\ldots,p_k) \nonumber \\
    &  & u^{\alpha_1\ldots\alpha_k}(p_1,\ldots,p_k)\,dp_1\ldots dp_k,
\end{eqnarray}
where
\begin{equation}
{{\mathcal M}^{\alpha_1^{\prime}\ldots\alpha_l^{\prime}}}_{\alpha_1\ldots\alpha_k}\in C^{\infty}(({\bf R}^4)^{k+l},{\bf C}),
\end{equation}
$\forall\alpha_1^{\prime}\ldots\alpha_l^{\prime},{\alpha_1\ldots\alpha_k}\in\{0,1,2,3\}$.

If the ${{\mathcal M}^{\alpha_1^{\prime}\ldots\alpha_l^{\prime}}}_{\alpha_1\ldots\alpha_k}$ functions are polynomially bounded as functions of $p_1^{\prime},\ldots,p_l^{\prime}, \newline p_1,\ldots,p_k$ to ${\bf C}$ then $\Xi=\Xi_{\mathcal M}$ is well defined as an mapping from ${\mathcal F}_k$ to ${\mathcal P}_l$.  
We would like to find the conditions under which such a $\Xi$ is $K$ intertwining. Let $\kappa\in K, u\in{\mathcal H}_k$. Then we have
\begin{eqnarray}
(\Xi(\kappa u))^{\alpha_1^{\prime}\ldots\alpha_l^{\prime}}(p_1^{\prime},\ldots,p_l^{\prime}) & = & \int {{\mathcal M}^{\alpha_1^{\prime}\ldots\alpha_l^{\prime}}}_{\alpha_1\ldots\alpha_k}(p_1^{\prime},\ldots,p_l^{\prime},p_1,\ldots,p_k) \nonumber \\
    &  & {\kappa^{\alpha_1}}_{\beta_1}\ldots{\kappa^{\alpha_k}}_{\beta_k}u^{\beta_1\ldots\beta_k}(\kappa^{-1}p_1,\ldots,\kappa^{-1}p_k)\,dp_1\ldots dp_k. \nonumber \\
 & = & \int {\kappa^{\alpha_1}}_{\beta_1}\ldots{\kappa^{\alpha_k}}_{\beta_k}{{\mathcal M}^{\alpha_1^{\prime}\ldots\alpha_l^{\prime}}}_{\alpha_1\ldots\alpha_k}(p_1^{\prime},\ldots,p_l^{\prime},\kappa p_1,\ldots,\kappa p_k) \nonumber \\
    &  & u^{\beta_1\ldots\beta_l}(p_1,\ldots,p_k)\,dp_1\ldots dp_k. \nonumber
\end{eqnarray}
Also
\begin{eqnarray}
(\kappa\Xi(u))^{\alpha_1^{\prime}\ldots\alpha_l^{\prime}}(p_1^{\prime},\ldots,p_l^{\prime}) & = & {\kappa^{\alpha_1^{\prime}}}_{\beta_1^{\prime}}\ldots{\kappa^{\alpha_l^{\prime}}}_{\beta_l^{\prime}}\Xi(u)^{\beta_1^{\prime}
\ldots\beta_l^{\prime}}(\kappa^{-1}p_1^{\prime},\ldots,\kappa^{-1}p_l^{\prime}, \nonumber \\
    & & p_1,\ldots,p_k) \nonumber \\
    & = & {\kappa^{\alpha_1^{\prime}}}_{\beta_1^{\prime}}\ldots{\kappa^{\alpha_l^{\prime}}}_{\beta_l^{\prime}}\int {{\mathcal M}^{\beta_1^{\prime}\ldots\beta_l^{\prime}}}_{\alpha_1\ldots\alpha_k}(\kappa^{-1}p_1^{\prime},\ldots,\kappa^{-1}p_l^{\prime}, \nonumber \\
    &  & p_1,\ldots,p_k)u^{\alpha_1\ldots\alpha_k}(p_1,\ldots,p_k)\,dp_1\ldots dp_k \nonumber \\
    & = & {\kappa^{\alpha_1^{\prime}}}_{\beta_1^{\prime}}\ldots{\kappa^{\alpha_l^{\prime}}}_{\beta_l^{\prime}}\int {{\mathcal M}^{\beta_1^{\prime}\ldots\beta_l^{\prime}}}_{\beta_1\ldots\beta_k}(\kappa^{-1}p_1^{\prime},\ldots,\kappa^{-1}p_l^{\prime}, \nonumber \\
    &  & p_1,\ldots,p_k)u^{\beta_1\ldots\beta_k}(p_1,\ldots,p_k)\,dp_1\ldots dp_k. \nonumber
\end{eqnarray}
It follows that $\Xi$ is intertwining if and only if
\begin{eqnarray}
    &  & {\kappa^{\alpha_1^{\prime}}}_{\beta_1^{\prime}}\ldots{\kappa^{\alpha_l^{\prime}}}_{\beta_l^{\prime}}{{\mathcal M}^{\beta_1^{\prime}\ldots\beta_l^{\prime}}}_{\beta_1\ldots\beta_k}(\kappa^{-1}p_1^{\prime},\ldots,\kappa^{-1}p_l^{\prime},p_1,\ldots,p_k) \nonumber \\
    & = & {\kappa^{\alpha_1}}_{\beta_1}\ldots{\kappa^{\alpha_k}}_{\beta_k}{{\mathcal M}^{\alpha_1^{\prime}\ldots\alpha_l^{\prime}}}_{\alpha_1\ldots\alpha_k}(p_1^{\prime},\ldots,p_l^{\prime},\kappa p_1,\ldots,\kappa p_k), \nonumber
\end{eqnarray}
for all $\kappa\in K; p_1^{\prime},\ldots,p_l^{\prime},p_1\ldots,p_k\in{\bf R}^4$ and all free indices. Therefore $\Xi$ is intertwining if and only if
\begin{eqnarray}
    &  & {\kappa^{\alpha_1^{\prime}}}_{\beta_1^{\prime}}\ldots{\kappa^{\alpha_l^{\prime}}}_{\beta_l^{\prime}}{{\mathcal M}^{\beta_1^{\prime}\ldots\beta_l^{\prime}}}_{\beta_1\ldots\beta_k}(p_1^{\prime},\ldots,p_l^{\prime},p_1,\ldots,p_k) \nonumber \\
    & = & {\kappa^{\alpha_1}}_{\beta_1}\ldots{\kappa^{\alpha_k}}_{\beta_k}{{\mathcal M}^{\alpha_1^{\prime}\ldots\alpha_l^{\prime}}}_{\alpha_1\ldots\alpha_k}(\kappa p_1^{\prime},\ldots,\kappa p_l^{\prime},\kappa p_1,\ldots,\kappa p_k), \label{eq:invariant_kernel}
\end{eqnarray}
for all $\kappa\in K; p_1^{\prime},\ldots,p_l^{\prime},p_1\ldots,p_k\in{\bf R}^4$ and all free indices. 

Now consider a more general class of operators $\Xi$ on ${\mathcal F}_k$ of the form 
\begin{eqnarray} \label{eq:general_kernel}
(\Xi(u))^{\alpha_1^{\prime}\ldots\alpha_l^{\prime}}(p_1^{\prime},\ldots,p_l^{\prime}) & = & \int {{\mathcal M}^{\alpha_1^{\prime}\ldots\alpha_l^{\prime}}}_{\alpha_1\ldots\alpha_k}(p_1^{\prime},\ldots,p_l^{\prime},\,dp_1,\ldots,\,dp_k) \nonumber \\
    &  & u^{\alpha_1\ldots\alpha_k}(p_1,\ldots,p_k),
\end{eqnarray}
where
\begin{equation}
{{\mathcal M}^{\alpha_1^{\prime}\ldots\alpha_l^{\prime}}}_{\alpha_1\ldots\alpha_k}:({\bf R}^4)^{l}\times{\mathcal B}_0({\bf R}^4)^k\rightarrow{\bf C},
\end{equation}
$\forall\alpha_1^{\prime}\ldots\alpha_l^{\prime},{\alpha_1\ldots\alpha_k}\in\{0,1,2,3\}$. The objects ${\mathcal M}$ defined in this fashion may be called tensor valued multikernels (comparing with the usual definition of kernel in the theory of stochastic processes (Revuz, 1984). We require that ${\mathcal M}(p_1^{\prime},\ldots,p_l^{\prime},\Lambda_1,\ldots,\Lambda_k)$ is smooth as a function of $p_i^{\prime}$ (all other arguments fixed) for each $i=1,\ldots,l$ and a tensor valued tempered measure as a function of $\Lambda_i$ (all other arguments fixed) for each $i=1,\ldots,k$. In this case $\Xi$ defined by Eq. \ref{eq:general_kernel} is a well defined operator on ${\mathcal H}_k$.

Arguing as above one can show that if ${\mathcal M}$ is a Lorentz invariant measure as a function of  each of its Borel set arguments (all other arguments held fixed) then $\Xi$ defined by Eq. \ref{eq:general_kernel} is intertwining if and only if
\begin{eqnarray}
    &  & {\kappa^{\alpha_1^{\prime}}}_{\beta_1^{\prime}}\ldots{\kappa^{\alpha_l^{\prime}}}_{\beta_l^{\prime}}{{\mathcal M}^{\beta_1^{\prime}\ldots\beta_l^{\prime}}}_{\beta_1\ldots\beta_k}(p_1^{\prime},\ldots,p_l^{\prime},\Gamma_1,\ldots,\Gamma_k) \nonumber \\
    & = & {\kappa^{\alpha_1}}_{\beta_1}\ldots{\kappa^{\alpha_k}}_{\beta_k}{{\mathcal M}^{\alpha_1^{\prime}\ldots\alpha_l^{\prime}}}_{\alpha_1\ldots\alpha_k}(\kappa p_1^{\prime},\ldots,\kappa p_l^{\prime},\kappa \Gamma_1,\ldots,\kappa \Gamma_k), 
\end{eqnarray}
for all $\kappa\in K; p_1^{\prime},\ldots,p_l^{\prime}\in{\bf R}^4,\Gamma_1,\ldots,\Gamma_k\in{\mathcal B}_0({\bf R}^4)$ and all free indices. 

\section{An example of a $K$ intertwining operator from ${\mathcal F}_2$ to ${\mathcal P}_2$}

To study 2 particle scattering we seek intertwining linear maps from ${\mathcal F}_2$ to ${\mathcal P}_l$. In particular, when the $|$out$>$ state contains 2 particles we are interested in the case when $l=2$. Let $m\in{\bf R}$. Define ${\mathcal M}_0$ by
\begin{equation} \label{eq:M_0_def}
{\mathcal M}_{0,\alpha_1^{\prime}\alpha_2^{\prime}\alpha_1\alpha_2}(p_1^{\prime},p_2^{\prime},p_1,p_2)  =  \overline{u}( p_1^{\prime},\alpha_1^{\prime})\gamma^{\rho}u(p_1,\alpha_1)\eta_{\rho\sigma}\overline{u}(p_2^{\prime},\alpha_2^{\prime})\gamma^{\sigma} u(p_2,\alpha_2). 
\end{equation}
where
\[ u(p,\alpha)=({\slas p}+m)e_{\alpha}, \mbox{ for }p\in{\bf R}^4, \alpha\in\{0,1,2,3\}, \]
and $\{e_{\alpha}:\alpha=0,1,2,3\}$ is the standard basis for  ${\bf C}^4$.
${\mathcal M}_0$ is a tensor valued order 2 polynomial finction of all its real arguments. Therefore it induces a linear mapping from ${\mathcal F}_2$ to ${\mathcal P}_2$.
\begin{theorem}
${{\mathcal M}_0^{\alpha_1^{\prime}\alpha_2^{\prime}}}_{\alpha_1\alpha_2}$ is an intertwining kernel.
\end{theorem}
{\bf Proof} Denote the mapping $p\mapsto{\slas p}$ by $\Sigma$. Then
\[ u(\kappa p,\alpha)=(\Sigma(\kappa p)+m)e_{\alpha}=(\kappa {\slas p}\kappa^{-1}+m)e_{\alpha}.\]
Therefore
\begin{eqnarray}
\overline{u}(\kappa p_1^{\prime},\alpha_1^{\prime})\gamma^{\rho}u(\kappa p_1,\alpha_1) & = & ((\kappa p_1^{\prime}\kappa^{-1}+m)e_{\alpha_1^{\prime}})^{\dagger}\gamma^0\gamma^{\rho}(\kappa{\slas p}\kappa^{-1}+m)e_{\alpha_1} \nonumber \\
    & = & e_{\alpha_1^{\dagger}}(\kappa^{-1\dagger}{\slas p_1^{\prime\dagger}}\kappa^{\dagger}+m)\gamma^0\gamma^{\rho}
(\kappa {\slas p_1}\kappa^{-1}+m)e_{\alpha_1} \nonumber
\end{eqnarray}
Now
\[ \kappa^{\dagger}\gamma^0\kappa=\gamma^0. \]
Therefore
\[ \kappa^{-1\dagger}=\kappa^{\dagger-1}=\gamma^0\kappa\gamma^0. \]
Hence
\begin{eqnarray}
\overline{u}(\kappa p_1^{\prime},\alpha_1^{\prime})\gamma^{\rho}u(\kappa p_1,\alpha_1) & = & e_{\alpha_1^{\prime}}^{\dagger}(\gamma^0\kappa\gamma^0\gamma^0{\slas p_1^{\prime}}\gamma^0\gamma^0\kappa^{-1}\gamma^0+m)\gamma^0\gamma^{\rho}(\kappa{\slas p_1}\kappa^{-1}+m)e_{\alpha_1} \nonumber \\
    & = & e_{\alpha_1^{\prime}}^{\dagger}(\gamma^0\kappa{\slas p_1^{\prime}}\kappa^{-1}\gamma^0+m)\gamma^0\gamma^{\rho}(\kappa{\slas p_1}\kappa^{-1}+m)e_{\alpha_1} \nonumber \\
    & = & e_{\alpha_1^{\prime}}^{\dagger}\gamma^0\kappa({\slas p_1^{\prime}}+m)\kappa^{-1}\gamma^{\rho}\kappa({\slas p_1}+m)\kappa^{-1}e_{\alpha_1} \nonumber \\
    & = & e_{\alpha_1^{\prime}}^{\dagger}\gamma^0\kappa({\slas p_1^{\prime}}+m){\Lambda^{-1\rho}}_{\mu}\gamma^{\mu}({\slas p_1}+m)\kappa^{-1}e_{\alpha_1} \nonumber
\end{eqnarray}
Similarly
\begin{equation}
\overline{u}(\kappa p_2^{\prime},\alpha_2^{\prime})\gamma^{\sigma}u(\kappa p_2,\alpha_2) =  e_{\alpha_2^{\prime}}^{\dagger}\gamma^0\kappa({\slas p_2^{\prime}}+m){\Lambda^{-1\sigma}}_{\nu}\gamma^{\nu}({\slas p_2}+m)\kappa^{-1}e_{\alpha_2} \nonumber
\end{equation}
It follows that
\begin{eqnarray}
{\mathcal M}_{0,\alpha_1^{\prime}\alpha_2^{\prime}\alpha_1\alpha_2}(\kappa p_1^{\prime},\kappa p_2^{\prime},\kappa p_1,\kappa p_2)  & = &  e_{\alpha_1^{\prime}}^{\dagger}\gamma^0\kappa({\slas p_1^{\prime}}+m)\gamma^{\mu}({\slas p_1}+m)\kappa^{-1}e_{\alpha_1}\eta_{\mu\nu} \nonumber \\
    &  & e_{\alpha_2^{\prime}}^{\dagger}\gamma^0\kappa({\slas p_2^{\prime}}+m)\gamma^{\nu} 
({\slas p_2}+m)\kappa^{-1}e_{\alpha_2} \nonumber
\end{eqnarray}
Now raise indices on ${\mathcal M}_0$ using the metric tensor $g=\gamma^0$.
\begin{eqnarray}
{{\mathcal M}_0^{\alpha_1^{\prime}\alpha_2^{\prime}}}_{\alpha_1\alpha_2}(\kappa p_1^{\prime},\kappa p_2^{\prime},\kappa p_1,\kappa p_2)  & = &  (\gamma^{0})^{\alpha_1^{\prime}\beta_1^{\prime}}(\gamma^0)^{\alpha_2^{\prime}\beta_2^{\prime}}{\mathcal M}_{0,\beta_1^{\prime}\beta_2^{\prime}\alpha_1\alpha_2}(\kappa p_1^{\prime},\kappa p_2^{\prime},\kappa p_1,\kappa p_2) \nonumber \\
    &  = & (\gamma^0)^{\alpha_1^{\prime}\beta_1^{\prime}}e_{\beta_1^{\prime}}^{\dagger}\gamma^0\kappa({\slas p_1^{\prime}}+m)\gamma^{\mu}({\slas p_1}+m)\kappa^{-1}e_{\alpha_1}\eta_{\mu\nu} \nonumber \\
    &  & (\gamma^0)^{\alpha_2^{\prime}\beta_2^{\prime}}e_{\beta_2^{\prime}}^{\dagger}\gamma^0\kappa({\slas p_2^{\prime}}+m)\gamma^{\nu}
({\slas p_2}+m)\kappa^{-1}e_{\alpha_2} \nonumber \\
    &  = & e_{\alpha_1^{\prime}}^{\dagger}\kappa({\slas p_1^{\prime}}+m)\gamma^{\mu}({\slas p_1}+m)\kappa^{-1}e_{\alpha_1}\eta_{\mu\nu} \nonumber \\
    &  & e_{\alpha_2^{\prime}}^{\dagger}\kappa({\slas p_2^{\prime}}+m)\gamma^{\nu}
({\slas p_2}+m)\kappa^{-1}e_{\alpha_2} \nonumber
\end{eqnarray}
Specializing to $\kappa=1_{K}$ we obtain that
\begin{eqnarray}
{{\mathcal M}_0^{\alpha_1^{\prime}\alpha_2^{\prime}}}_{\alpha_1\alpha_2}(p_1^{\prime},p_2^{\prime},p_1, p_2) & = &
 e_{\alpha_1^{\prime}}^{\dagger}({\slas p_1^{\prime}}+m)\gamma^{\mu}({\slas p_1}+m)e_{\alpha_1}\eta_{\mu\nu} \nonumber \\
    &  & e_{\alpha_2^{\prime}}^{\dagger}({\slas p_2^{\prime}}+m)\gamma^{\nu}({\slas p_2}+m)e_{\alpha_2} \nonumber 
\end{eqnarray}
Now
\begin{eqnarray}
e_{\alpha_1^{\prime}}^{\dagger}\kappa({\slas p_1^{\prime}}+m)\gamma^{\mu}({\slas p_1}+m)\kappa^{-1}e_{\alpha_1} & = & 
{[\kappa({\slas p_1^{\prime}}+m)\gamma^{\mu}({\slas p_1}+m)\kappa^{-1}]^{\alpha_1^{\prime}}}_{\alpha_1} \nonumber \\
    & = & {\kappa^{\alpha_1^{\prime}}}_{\beta_1^{\prime}}{{[({\slas p_1^{\prime}}+m)\gamma^{\mu}({\slas p_1}+m)]}^{\beta_1^{\prime}}}_{\rho_1}{(\kappa^{-1})^{\rho_1}}_{\alpha_1} \nonumber \\
    &  = & {\kappa^{\alpha_1^{\prime}}}_{\beta_1^{\prime}}{{(\kappa^{-1})}^{\rho_1}}_{\alpha_1}           e_{\beta_1^{\prime}}^{\dagger}({\slas p_1^{\prime}}+m)\gamma^{\mu}({\slas p_1}+m)e_{\rho_1} \nonumber
\end{eqnarray}
Similarly
\begin{eqnarray}
e_{\alpha_2^{\prime}}^{\dagger}\kappa({\slas p_2^{\prime}}+m)\gamma^{\nu}({\slas p_2}+m)\kappa^{-1}e_{\alpha_2} & = & 
    {\kappa^{\alpha_2^{\prime}}}_{\beta_2^{\prime}}{(\kappa^{-1})^{\rho_2}}_{\alpha_2}e_{\beta_2^{\prime}}^{\dagger}({\slas p_2^{\prime}}+m)\gamma^{\nu}({\slas p_2}+m)e_{\rho_2} \nonumber
\end{eqnarray}
Therefore
\[ {{\mathcal M}_0^{\alpha _1^{\prime}\alpha_2^{\prime}}}_{\alpha_1\alpha_2}(\kappa p_1^{\prime},\kappa p_2^{\prime},\kappa p_1,\kappa p_2) \]  
\[ = {\kappa^{\alpha_1^{\prime}}}_{\beta_1^{\prime}}{(\kappa^{-1})^{\beta_1}}_{\alpha_1} {\kappa^{\alpha_2^{\prime}}}_{\beta_2^{\prime}}{(\kappa^{-1})^{\beta_2}}_{\alpha_2}{{\mathcal M}_0^{\beta_1^{\prime}\beta_2^{\prime}}}_{\beta_1\beta_2}(p_1^{\prime},p_2^{\prime},p_1, p_2) \]
for all $\kappa\in K, p_1^{\prime},p_2^{\prime},p_1,p_2\in{\bf R}^4$ and all free indices, which is equivalent to the condition that ${\mathcal M}_0$ be intertwining. $\Box$

\section{Fermions, bosons and the supersymmetry algebra}

A kernel is a map ${\mathcal M}:\{0,1,2,3\}^k\times\{0,1,2,3\}^l\times({\bf R}^4)^k\times({\bf R^4)}^l\rightarrow {\bf C}$. We assume that kernels ${\mathcal M}$ are at least Borel functions of their real arguments. A kernel will be said to be Borel/continuous/smooth if it is a Borel/continuous/smooth function of all its real arguments. For the purposes of this paper we will consider kernels that are $K$ intertwining, $C^{\infty}$ and polynomially bounded. Let ${\mathcal K}_{kl}$ denote the space of all such kernels of size $(k,l)$. Then
\begin{eqnarray}
{\mathcal M}\in{\mathcal K}_{kl} & \Leftrightarrow & ({{\mathcal M}^{\alpha}}_{\beta}\in C^{\infty}(({\bf R}^4)^k\times({\bf R}^4)^l,{\bf C}),\forall\alpha\in\{0,1,2,3\}^k,\beta\in\{0,1,2,3\}^l \nonumber \\
    &  & \mbox{ and ${\mathcal M}$ is $K$ intertwining and polynomially bounded}). \nonumber
\end{eqnarray}

Given a linear map $\Xi:{\mathcal H}_k\rightarrow{\mathcal D}_l$ there is induced an element $T_{\Xi}\in\mbox{Alt}_k({\mathcal H}_1,{\mathcal D})$ defined by
\begin{equation}
T_{\Xi}(u_1,\ldots,u_k)=\Xi(u_1\wedge\ldots\wedge u_k).
\end{equation}
If $\Xi$ is $K$ intertwining then, since $K$ acts on the Grassmann algebra ${\mathcal H}$ by algebra automorphisms, and by definition of the action of $K$ on Alt$({\mathcal H}_1,{\mathcal D})$,
\begin{eqnarray}
(\kappa T_{\Xi})(v_1,\ldots,v_n) & = & \kappa T_{\Xi}(\kappa^{-1}u_1,\ldots,\kappa^{-1}u_k) \nonumber \\
    & = & \kappa\Xi(\kappa^{-1}u_1\wedge\ldots\wedge\kappa^{-1}u_k) \nonumber \\
    & = & \kappa\Xi(\kappa^{-1}(u_1\wedge\ldots\wedge u_k)) \nonumber \\
    & = & \kappa\kappa^{-1}\Xi(u_1\wedge\ldots\wedge u_k) \nonumber \\
    & = & \Xi(u_1\wedge\ldots\wedge u_k) \nonumber \\
    & = & T_{\Xi}(u_1,\ldots,u_k), \nonumber 
\end{eqnarray}
for all $\kappa\in K,u_1,\ldots,u_k\in{\mathcal H}_1$.
Thus $T_{\Xi}\in{\mathcal J}_k({\mathcal H}_1,{\mathcal D})$. Therefore all the machinery of supersymmetry described in Section~\ref{section:superfields} applies.

\section{The kernel algebra}

Let 
\begin{equation}
{\mathcal K}=\bigoplus_{k,l=0}^{\infty}{\mathcal K}_{kl}.
\end{equation}
Then ${\mathcal K}$ forms a doubly graded algebra under the operation $({\mathcal M}_1,{\mathcal M}_2)\mapsto{\mathcal M}_1\otimes{\mathcal M}_2$ defined by
\begin{equation}
({{\mathcal M}_1\otimes{\mathcal M}_2)^{\alpha_1^{\prime}\ldots\alpha_{l}^{\prime}}}_{\alpha_1\ldots\alpha_{k}}(p_1^{\prime},\ldots, p_{l}^{\prime},p_1,\ldots,p_{k})
\end{equation} 
\[ ={{{\mathcal M}_1}^{\alpha_1^{\prime}\ldots\alpha_{l_1}^{\prime}}}_{\alpha_1\ldots\alpha_{k_1}}(p_1^{\prime},\ldots,p_{l_1}^{\prime},p_1,\ldots,p_{k_1}) \]
\[ {{\mathcal M}_2^{\alpha_{(l_1+1)}^{\prime}\ldots\alpha_{l}^{\prime}}}_{\alpha_{(k_1+1)}\ldots\alpha_{k}}(p_{(l_1+1)}^{\prime},\ldots,p_{l}^{\prime},p_{(k_1+1)},\ldots,p_{k}), \]
for ${\mathcal M}_1\in{\mathcal K}_{k_1l_1}$, ${\mathcal M}_2\in{\mathcal K}_{k_2l_2}, k_1,k_2,l_1,l_2\in\{0,1,2,\ldots\}, k=k_1+k_2, l=l_1+l_2$.

By a straightforward calculation one can show that 
\begin{equation}
\Xi_{{\mathcal M}_1\otimes{\mathcal M}_2}(u_1\otimes u_2)=\Xi_{{\mathcal M}_1}(u_1)\otimes\Xi_{{\mathcal M}_2}(u_2), 
\end{equation}
where $\Xi_{\mathcal M}$ is the scattering operator defined by the kernel ${\mathcal M}$.

\section{Fermion loops}

We will now consider operators associated with fermion loops. Consider the expression
\begin{equation}
\Pi^{\mu\nu}(q)=\int\gamma^{\mu}S_F(p)\gamma^{\nu}S_F(p-q)\,dp,
\end{equation}
(see (Itzikson and Zuber, 1980, p. 319)). $\Pi^{\mu\nu}$ has the form of a convolution of fermion propagators, which are themselves matrix valued measures therefore $\Pi^{\mu\nu}$ is formally a matrix valued measure. We can formally compute, for $\psi\in{\cal S}({\bf R}^4,{\bf C})$, 
\begin{eqnarray}
<\Pi^{\mu\nu},\psi> & = & \int\Pi^{\mu\nu}(q)\psi(q)\,dq \nonumber \\
    &  = & \int\!\!\int\gamma^{\mu}S_F(p)\gamma^{\nu}S_F(p-q)\psi(q)\,dp\,dq \nonumber \\
    & = & \int\!\!\int\gamma^{\mu}S_F(p)\gamma^{\nu}S_F(p-q)\psi(q)\,dq\,dp \nonumber \\
    & = & \int\!\!\int\gamma^{\mu}S_F(p)\gamma^{\nu}S_F(q)\psi(p-q)\,dq\,dp \nonumber \\
    & = & \int\gamma^{\mu}S_F(p)\gamma^{\nu}\int S_F(q)\psi(p-q)\,dq\,dp \nonumber \\
    & = & \int\gamma^{\mu}({\slas p}+m)\gamma^{\nu}\int({\slas q}+m)\psi(p-q)\,\Omega_m(dq)\,\Omega_{-m}(dp) \nonumber \\
   & = & \int\!\!\int\gamma^{\mu}({\slas p}+m)\gamma^{\nu}({\slas q}+m)\psi(p-q)\,\Omega_{-m}(dp)\,\Omega_{m}(dq) \nonumber
\end{eqnarray}
where we have assumed that the particles associated with the loop are a fermion/anti-fermion pair.
Note that $\Pi^{\mu\nu}$ defined by
\begin{equation}
<\Pi^{\mu\nu},\psi> =   \int\!\!\int\gamma^{\mu}({\slas p}+m)\gamma^{\nu}({\slas q}+m)\psi(p-q)\,\Omega_{-m}(dp)\,\Omega_{m}(dq),
\end{equation}
is a well defined and continuous ${\bf C}^{4\times 4}$ valued operator on ${\cal S}({\bf R}^4,{\bf C})$ (the integral exists because $|p-q|\rightarrow\infty$ as $|p|, |q|\rightarrow\infty$ on their mass shells). 

$\Pi^{\mu\nu}$ clearly extends to include the space of all measurable $\psi:{\bf R}^4\rightarrow{\bf C}$ which are bounded and have compact support. Therefore, evaluating it on characteristic functions $\chi_{\Gamma}$ of the form 
\[ \psi(p)=\chi_{\Gamma}(p), \Gamma\in{\mathcal B}_0({\bf R}^4), \]
where 
\begin{equation}
\chi_{\Gamma}(p)=\left\{\begin{array}{l}
1 \mbox{ if } p\in\Gamma \nonumber \\
0 \mbox{ otherwise},
\end{array}\right.
\end{equation}
we deduce that $\Pi^{\mu\nu}$ is a (Borel) ${\bf C}^{4\times4}$ matrix valued tempered measure on ${\bf R}^4$. 

Define $\Pi=\eta_{\mu\nu}\Pi^{\mu\nu}.$
\begin{theorem}
$\Pi$ is a $K$ covariant measure.
\end{theorem}
{\bf Proof}
\begin{eqnarray}
<\Pi,\kappa \psi> & = & \int\!\!\int\eta_{\mu\nu}\gamma^{\mu}({\slas p}+m)\gamma^{\nu}({\slas q}+m)(\kappa \psi)(p-q)\,\Omega_{-m}(dp)\,\Omega_{m}(dq) \nonumber \\
     & = & \int\!\!\int\eta_{\mu\nu}\gamma^{\mu}({\slas p}+m)\gamma^{\nu}({\slas q}+m)\psi(\Lambda^{-1}(p-q))\,\Omega_{-m}(dp)\,\Omega_{m}(dq) \nonumber \\
     & = & \int\!\!\int\eta_{\mu\nu}\gamma^{\mu}(\Sigma(\Lambda p)+m)\gamma^{\nu}(\Sigma(\Lambda q)+m) \nonumber \\
    & & \psi(p-q)\,\Omega_{-m}(dp)\,\Omega_{m}(dq) \nonumber \\
 & = & \int\!\!\int\eta_{\mu\nu}\gamma^{\mu}\kappa({\slas p}+m)\kappa^{-1}\gamma^{\nu}\kappa({\slas q}+m)\kappa^{-1} \psi(p-q)\,\Omega_{-m}(dp)\,\Omega_{m}(dq) \nonumber \\
 & = & \kappa\int\!\!\int\eta_{\mu\nu}\kappa^{-1}\gamma^{\mu}\kappa({\slas p}+m)\kappa^{-1}\gamma^{\nu}\kappa({\slas q}+m)\psi(p-q)\,\Omega_{-m}(dp) \nonumber \\
    &  & \,\Omega_{m}(dq)\kappa^{-1} \nonumber \\
 & = & \kappa\int\!\!\int\eta_{\mu\nu}{\Lambda^{-1\mu}}_{\rho}\gamma^{\rho}({\slas p}+m){\Lambda^{-1\nu}}_{\sigma}\gamma^{\sigma}({\slas q}+m)\psi(p-q) \nonumber \\
    &  &  \,\Omega_{-m}(dp)\,\Omega_{m}(dq)\kappa^{-1} \nonumber \\
   & = & \kappa\int\!\!\int\eta_{\rho\sigma}\gamma^{\rho}({\slas p}+m)\gamma^{\sigma}({\slas q}+m)\psi(p-q)\,\Omega_{-m}(dp)\,\Omega_{m}(dq)\kappa^{-1} \nonumber \\
    & = & \kappa<\Pi,\psi>\kappa^{-1}. \nonumber
\end{eqnarray}
as required. $\Box$

Here we have used the fundamental intertwining property of the Feynman slash operator (see Mashford, 2017), the Lorentz invariance of the measure $\Omega_m$ and the defining characteristic of the Lorentz transformations. We have use the symbol $\Sigma$ to denote the map $p \mapsto{\slas p}$. Thus a fermion loop consisting of 2 fermion propagators is associated with an intertwining matrix valued measure $\Pi$. It is clear how this construction generalizes to fermion loops consisting of $n$ fermion or anti-fermion propagators. 

It remains to be shown that QFT computations using $\Pi$ agree with the usual computations of QFT using renormalization. This will be addressed in a subsequent paper.

\section{The electron self energy}
We now consider the operator associated with the fermion self-energy. Consider the expression
\begin{equation}
\Gamma(p) = \int D_{F\mu\nu}(q)\gamma^{\mu}S_F(p-q)\gamma^{\nu}\,dq,
\end{equation}
in which $D_{F\mu\nu}$ is the photon propagator and $S_F$ is the fermion propagator (see (Itzikson and Zuber, 1980, p. 329). $\Gamma$ has the form of a convolution of a complex measure ($D_{F\mu\nu}$) and a matrix valued measure ($S_F$) and therefore formally $\Gamma$ is a matrix valued measure. We may formally compute, for $\psi\in{\cal S}({\bf R}^4,{\bf C})$
\begin{eqnarray}
<\Gamma,\psi> & = & \int\Gamma(p)\psi(p)\,dp \nonumber \\
    & = & \int\!\!\int D_{F\mu\nu}(q)\gamma^{\mu}S_F(p-q)\gamma^{\nu}\psi(p)\,dq\,dp \nonumber \\
    & = &  \int\!\!\int D_{F\mu\nu}(q)\gamma^{\mu}S_F(p-q)\gamma^{\nu}\psi(p)\,dp\,dq \nonumber \\
    & = &  \int\!\!\int D_{F\mu\nu}(q)\gamma^{\mu}S_F(p)\gamma^{\nu}\psi(p+q)\,dp\,dq \nonumber \\
    & = &  \int\!\!\int\eta_{\mu\nu}\gamma^{\mu}({\slas p}+m)\gamma^{\nu}\psi(p+q)\,\Omega_m(dp)\,\Omega_0^{+}(dq) \nonumber \\
\end{eqnarray}
Therefore, writing
\begin{equation}
<\Gamma,\psi>=\int\!\!\int\eta_{\mu\nu}\gamma^{\mu}({\slas p}+m)\gamma^{\nu}\psi(p+q)\,\Omega_m(dp)\,\Omega_0^{+}(dq),
\end{equation}
we may consider $\Gamma$ to be a map from ${\cal S}({\bf R}^4,{\bf C})$ to ${\bf C}^{4\times 4}$.  
Note that the $\Gamma$ is a well defined and continuous operator. Itzikson and Zuber (1980, p. 329) write, in regards to the electron self-energy, ``This expression, a $4\times 4$ matrix function of $p$, unfortunately suffers from all possible diseases" and then go on to discuss the nature of its divergences. We, using a formal argument and a rigorous formulation of the propagators, have given it a well defined meaning as a matrix valued measure.
\begin{theorem}
$\Gamma$ is a $K$ covariant measure.
\end{theorem}
{\bf Proof}
\begin{eqnarray}
<\Gamma,\kappa \psi> & = & \int\!\!\int\eta_{\mu\nu}\gamma^{\mu}({\slas p}+m)\gamma^{\nu}(\kappa \psi)(p+q)\,\Omega_0^{+}(dq)\,\Omega_m(dp) \nonumber \\
     & = & \int\!\!\int\eta_{\mu\nu}\gamma^{\mu}({\slas p}+m)\gamma^{\nu}\psi(\Lambda^{-1}(p+q))\,\Omega_0^{+}(dq)\,\Omega_m(dp) \nonumber \\
    & = & \int\!\!\int\eta_{\mu\nu}\gamma^{\mu}(\Sigma(\Lambda p))+m)\gamma^{\nu}\psi(p+q)\Omega_0^{+}(dq)\Omega_m(dp) \nonumber \\
    & = & \int\!\!\int\eta_{\mu\nu}\gamma^{\mu}(\kappa{\slas p}\kappa^{-1}+m)\gamma^{\nu}\psi(p+q)\,\Omega_0^{+}(dq)\,\Omega_m(dp) \nonumber \\
    & = & \int\!\!\int\eta_{\mu\nu}\kappa{\Lambda^{-1\mu}}_{\rho}\gamma^{\rho}({\slas p}+m){\Lambda^{-1\nu}}_{\sigma}\gamma^{\sigma}\kappa^{-1}\psi(p+q)\, \nonumber \\
    & & \Omega_0^{+}(dq)\,\Omega_m(dp) \nonumber \\
    & = & \kappa\int\!\!\int\eta_{\rho\sigma}\gamma^{\rho}({\slas p}+m)\gamma^{\sigma}\psi(p+q)\,\Omega_0^{+}(dq)\,\Omega_m(dp)\kappa^{-1} \nonumber \\
    & = & \kappa<\Gamma,\psi>\kappa^{-1} \nonumber
\end{eqnarray}
as required. $\Box$

It remains to be shown that QFT computations using $\Gamma$ agree with the usual computations of QFT using renormalization. This will be addressed in a subsequent paper.

\section{Quantum electrodynamics (QED) \label {section:QED}}

\subsection{$K$ covariant scalar valued kernels}

A map $L:{\bf R}^4\times{\mathcal B}_0({\bf R}^4)\rightarrow{\bf C}$ will be said to be a (scalar valued) kernel if $\Gamma\mapsto L(p,\Gamma)$ is a complex measure on ${\bf R}^4$ for each $p\in{\bf R}^4$. $L$ will be said to be measurable/continuous/smooth if $p\mapsto L(p,\Gamma)$ is measurable/continuous/smooth for all $\Gamma\in{\mathcal B}_0({\bf R}^4)$ (compare with Revuz, 1984). $L$ will be said to be tempered if $\Gamma\mapsto L(p,\Gamma)$ is a tempered measure for all $p\in{\bf R}^4$.

A scalar valued kernel $L$ will be said to be $K$ covariant if 
\begin{equation}
L(\kappa p,\kappa\Gamma)=L(p,\Gamma), \forall \kappa\in K, p\in{\bf R}^4, \Gamma\in{\mathcal B}_0({\bf R}^4).
\end{equation}

Suppose that $\mu:{\mathcal B}_0({\bf R}^4)\rightarrow{\bf C}$ is a $K$ invariant complex measure. Define $L=L_{\mu}:{\bf R}^4\times{\mathcal B}_0({\bf R}^4)\rightarrow{\bf C}$ by
\begin{equation}
L(p,\Gamma)=\mu(\Gamma-p).
\end{equation}
Then $L$ is readily seen to be a $K$ covariant kernel and will be called the standard kernel associated with the measure $\mu$.

\subsection{A useful theorem}

Recalling the notation of Section \ref{Feynman} let
\begin{eqnarray}
{\mathcal X}^{+} & = & ({\mathcal S}^{+})^{\wedge} \\
    & = & \{{\chp \psi}:\psi\in{\mathcal S}({\bf R}^4,{\bf C})\mbox{ is such that }\psi(x)=0, \mbox{ for all }x\mbox{ for which }x^0<0\}, \nonumber
\end{eqnarray}
be the space of positive energy Schwartz wave packets and, similarly, ${\mathcal X}^{-} = ({\mathcal S}^{-})^{\wedge}$. 
Using the techniques of Section \ref{Feynman} we can prove the following rigorous (not just formal) result.
\begin{theorem} \label{theorem:propagator}
Let $\psi\in {\mathcal S}^{+}$ and $m\ge 0$. Then
\begin{equation}
-\pi i\int{\chp \psi}(p)\,\Omega_m^{+}(dp)=\int\frac{{\chp \psi}(p)}{(p^2-m^2)+i\epsilon}\,dp.
\end{equation}
\end{theorem}
{\bf Proof}
\begin{eqnarray}
-\pi i\int{\chp \psi}(p)\,\Omega_m^{+}(dp) & = & -\pi i\int_{{\bf R}^3}{\chp \psi}(\omega_m({\vct p}),{\vct p})\,\frac{d{\vct p}}{\omega_m({\vct p})} \nonumber \\
     & = & -\pi i\lim_{R\rightarrow\infty}\int_{B_R}{\chp \psi}(\omega_m({\vct p}),{\vct p})\,\frac{d{\vct p}}{\omega_m({\vct p})} \nonumber \\
    & = & -\pi i\lim_{R\rightarrow\infty}\int_{B_R}(2\pi)^{-2}\int\psi(x)\frac{e^{-i(\omega_m({\vct p})x^0-{\vct p}.{\vct x})}}{\omega_m({\vct p})}\,dx\,d{\vct p} \nonumber \\
    & = & \lim_{R\rightarrow\infty}(2\pi)^{-2}\int_{S^{+}}\int_{B_R}\lim_{\eta\rightarrow 0^{+}}\int_{\bf R}\frac{e^{-ip.x}}{(p^0)^2-(\omega_m({\vct p})-i\eta)^2}\,dp^0 \nonumber \\
    &  & d{\vct p}\,\psi(x)\,dx \nonumber \\
    & = & \lim_{R\rightarrow\infty}\int_{B_R}\lim_{\eta\rightarrow 0^{+}}\int_{\bf R}\frac{{\chp \psi}(p)}{(p^0)^2-(\omega_m({\vct p})-i\eta)^2}\,dp^0\,d{\vct p} \nonumber \\
    & = & \int_{{\bf R}^3}\lim_{\eta\rightarrow 0^{+}}\int_{\bf R}\frac{{\chp \psi}(p)}{(p^0)^2-(\omega_m({\vct p})-i\eta)^2}\,dp^0\,d{\vct p} \nonumber \\
    & = & \int\frac{{\chp \psi}(p)}{p^2-m^2+i\epsilon}\,dp\nonumber 
 \end{eqnarray} 
in which $B_R=\{{\vct p}\in{\bf R}^3:|{\vct p}|<R\}$. Here we have used the definition of the $i$-epsilon procedure, the dominated convergence theorem, Fubini's theorem and the Cauchy residue theorem. $\Box$

From this theorem it follows that
\begin{equation}
\int\psi(p)\,\Omega_m^{+}(dp)=\frac{i}{\pi}\int\frac{\psi(p)}{p^2-m^2+i\epsilon}\,dp, \forall\psi\in{\mathcal X}^{+}, m\ge 0.
\end{equation}

\subsection{A general construction for generating S matrices \label{section:general_construction}}

Suppose that ${\mathcal N}\in{\mathcal K}_{22}$ and $L$ is a $K$ covariant tempered scalar kernel. We can construct a $K$ intertwining operator $\Xi$ using ${\mathcal N}$ and $L$ as follows.
Define the operator $\Xi$ by
\begin{equation} \label{Moller_kernel}
(\Xi(u))^{\alpha_1^{\prime}\alpha_2^{\prime}}(p_1^{\prime},p_2^{\prime})=\int{{{\mathcal N}^{\alpha_1^{\prime}\alpha_2^{\prime}}}_{\alpha_1\alpha_2}(p_1^{\prime},p_2^{\prime}},p_1,p_2)u^{\alpha_1\alpha_2}(p_1,p_2)\,dp_1L(p_2^{\prime},dp_2),
\end{equation}
where $u\in{\mathcal H}_2,p_1^{\prime},p_2^{\prime}\in{\bf R}^4,{\alpha_1^{\prime}},{\alpha_2^{\prime}}\in\{0,1,2,3\}.$
Then it is fairly easy to show that $\Xi$ is a $K$ intertwining operator. We have given this construction in the case when the $|$in$>$ states and the $|$out$>$ states both have 2 particles, i.e. $k=l=2$. The generalization to arbitrary $k$ and $l$ is straightforward. 

This construction is a special case of a more general construction for generating S matrices from given kernels using the algebra of kernels, which will be discussed more fully in a subsequent paper.

\subsection{Fermion-fermion scattering to the first order}

A very simple covariant scalar valued kernel to consider is kernel $L$ induced by the measure $\Omega_0^{+}$. It is straightforward to show that
\begin{equation}
\int\psi(p)L(p^{\prime},dp)=\int\psi(p+p^{\prime})\,\Omega_0^{+}(dp), \forall\psi\in{\mathcal S}({\bf R}^4,{\bf C}), p^{\prime}\in{\bf R}^4.
\end{equation}

As a result of Theorem \ref{theorem:propagator} we have that
\begin{equation}
\int\psi(p+p^{\prime})\,\Omega_0^{+}(dp)=\frac{i}{\pi}\int\frac{\psi(p)}{(p-p^{\prime})^2+i\epsilon}\,dp, \forall\psi\in{\mathcal X}^{+}, p^{\prime}\in{\bf R}^4.
\end{equation}
Also it is straightforward to show that if $\psi\in{\mathcal X}^{+}$ and $P:{\bf R}^4\rightarrow{\bf C}$ is a polynomial function then $P\psi\in{\mathcal X}^{+}$.
Therefore, since ${\mathcal M}_0$ is a polynomial function of all its arguments, the operator generated by the kernel ${\mathcal M}_0$ and $L$ according to the construction of Section \ref{section:general_construction} is given by\begin{equation} \label{Moller_kernel}
(\Xi(u))^{\alpha_1^{\prime}\alpha_2^{\prime}}(p_1^{\prime},p_2^{\prime})=\frac{i}{\pi}\int{{{\mathcal M}^{\alpha_1^{\prime}\alpha_2^{\prime}}}_{\alpha_1\alpha_2}(p_1^{\prime},p_2^{\prime}},p_1,p_2)u^{\alpha_1\alpha_2}(p_1,p_2)\,dp_1\,dp_2.
\end{equation}
where
\begin{equation}
{{\mathcal M}^{\alpha_1^{\prime}\alpha_2^{\prime}}}_{\alpha_1\alpha_2}(p_1^{\prime},p_2^{\prime},p_1,p_2)=\frac{{{\mathcal M_0}^{\alpha_1^{\prime}\alpha_2^{\prime}}}_{\alpha_1\alpha_2}(p_1^{\prime},p_2^{\prime},p_1,p_2)}{(p_2-p_2^{\prime})^2+i\epsilon}
\end{equation}
We are interested in finding $K$ intertwining operators which map to antisymmetric fermionic particle states. Therefore we antisymmetrize the operator $\Xi$ defined above resulting in the operator $\Xi$ defined by
\begin{eqnarray}
(\Xi(u))^{\alpha_1^{\prime}\alpha_2^{\prime}}(p_1^{\prime},p_2^{\prime}) & = & \frac{i}{2\pi}\int{{\mathcal M}^{\alpha_1^{\prime}\alpha_2^{\prime}}}_{\alpha_1\alpha_2}(p_1^{\prime},p_2^{\prime},p_1,p_2)u^{\alpha_1\alpha_2}(p_1,p_2)\,dp_1\,dp_2 \nonumber \\
    & = &  \frac{i}{2\pi}\int(\gamma^0)^{\alpha_1^{\prime}\beta_1^{\prime}}(\gamma^0)^{\alpha_2^{\prime}\beta_2^{\prime}}{\mathcal M}_{\beta_1^{\prime}\beta_2^{\prime}\alpha_1\alpha_2}(p_1^{\prime},p_2^{\prime},p_1,p_2) \nonumber \\
    &  & u^{\alpha_1\alpha_2}(p_1,p_2)\,dp_1\,dp_2, \nonumber
\end{eqnarray}
where 
\begin{equation} \label{eq:Moller_Feynman_amplitude}
{\mathcal M}_{\beta_1^{\prime}\beta_2^{\prime}\alpha_1\alpha_2}(p_1^{\prime},p_2^{\prime},p_1,p_2)=\frac{{\mathcal M}_{0,\beta_1^{\prime}\beta_2^{\prime}\alpha_1\alpha_2}(p_1^{\prime},p_2^{\prime},p_1,p_2)}{(p_2-p_2^{\prime})^2+i\epsilon}- \frac{{\mathcal M}_{0,\beta_2^{\prime}\beta_1^{\prime}\alpha_1\alpha_2}(p_2^{\prime},p_1^{\prime},p_1,p_2)}{(p_2-p_1^{\prime})^2+i\epsilon},
\end{equation}
in which ${\mathcal M}_0$ is given by Eq. \ref{eq:M_0_def}. We recognize in Eq. \ref{eq:Moller_Feynman_amplitude} the Feynman amplitude for M\o ller electron-electron scattering (Itzikson and Zuber, 1980, p. 278) which leads to well verified experimental predictions.

\section{The weak interaction}

\subsection{States and kernels with Lorentz indices}

One can consider kernels with Lorentz indices as well as $K$ indices as follows. We will denote $K$-indices by $\alpha_1,\beta_1,\alpha_2,\beta_2,\ldots$ and primed versions of these and Lorentz indices by $\mu_1,\nu_1,\mu_2,\nu_2,\ldots$ and primed versions of these. Then we define a kernel of type $(k^{\prime},l^{\prime},k,l)$ i.e. with $k^{\prime}$ contravariant Lorentz indices, $l^{\prime}$ contravariant $K$-indices, $k$ covariant Lorentz indices and $l$ covariant $K$-indices to be a map ${\mathcal M}:\{0,1,2,3\}^{4(k^{\prime}+l^{\prime}+k+l)}\times({\bf R}^4)^{l^\prime}\times({\mathcal B}_0({\bf R}^4))^l\rightarrow{\bf C}$ which is smooth and polynomially bounded in its continuous arguments and a tempered measure in its set arguments.

Let ${\mathcal F}_{k_1k_2}$ (${\mathcal P}_{k_1k_2}$) denote the space of smooth tensor valued Schwartz (polynomially bounded) functions with $k_1$ $K$ indices and $k_2$ Lorentz indices. Let ${\mathcal K}_{k_1k_2l_1l_2}$ denote the set of kernels mapping ${\mathcal F}_{k_1k_2}$ to ${\mathcal P}_{l_1l_2}$.

A kernel will be said to transform covariantly if
\begin{align*}
    &  {{\mathcal M}^{{\mu^{\prime}_1}\ldots\mu_{k^{\prime}}^{\prime}\alpha_1^{\prime}\ldots\alpha_{l^{\prime}}^{\prime}}}_{\mu_1\ldots\mu_{k}\alpha_1\ldots\alpha_l}(\kappa p_1^{\prime},\ldots,\kappa p_{l^{\prime}}^{\prime},\kappa\Gamma_1,\ldots,\kappa\Gamma_l) \\
    & = {(\Lambda^{-1})^{\mu_1^{\prime}}}_{\nu_1^{\prime}}\ldots{{(\Lambda^{-1})}^{\mu^{\prime}_{k^{\prime}}}}_{\nu^{\prime}_{k^{\prime}}}{\Lambda^{\nu_1}}_{\mu_1}\ldots{\Lambda^{\nu_k}}_{\mu_k}   \\
    &  {\kappa^{\alpha_1^{\prime}}}_{\beta_1^{\prime}}\ldots{\kappa^{\alpha^{\prime}_{l^{\prime}}}}_{\beta^{\prime}_{l^{\prime}}}{(\kappa^{-1})^{\beta_1}}_{\alpha_1}\ldots{{(\kappa^{-1})}^{\beta_l}}_{\alpha_l} \\
    & {{\mathcal M}^{\nu_1^{\prime}\ldots\nu_{k^{\prime}}^{\prime}\beta_1^{\prime}\ldots\beta_{l^{\prime}}^{\prime}}}_{\nu_1\ldots\nu_k\beta_1\ldots\beta_l}(p_1^{\prime},\ldots,p_{l^{\prime}}^{\prime},\Gamma_1,\ldots,\Gamma_l),
\end{align*}
for all $\kappa\in K,p_1^{\prime},\ldots,p_{l^{\prime}}^{\prime}\in{\bf R}^4,\Gamma_1,\ldots,\Gamma_l\in{\mathcal B}_0({\bf R}^4)$ and all free indices, where $\Lambda=\Lambda(\kappa)$.
\begin{theorem}
The W boson propagator (in momentum space) transforms covariantly.
\end{theorem}
{\bf Proof}\\ 
\begin{align*}
D_{m_W}^{\mu_1\mu_2}(\kappa\Gamma) & = -\pi i\int_{\kappa\Gamma}(\eta^{\mu_1\mu_2}-m_W^{-2}q^{\mu_1}q^{\mu_2})\,\Omega_{m_W}(dq) \\
    & = -\pi i\int_{\Gamma}(\eta^{\mu_1\mu_2}-m_W^{-2}(\Lambda q)^{\mu_1}(\Lambda q)^{\mu_2})\,\Omega_{m_W}(dq) \\
    & = -\pi i\int_{\Gamma}(\eta^{\mu_1\mu_2}-{\Lambda^{\mu_1}}_{\nu_1}{\Lambda^{\mu_2}}_{\nu_2}m_W^{-2}q^{\nu_1}q^{\nu_2})\,\Omega_{m_W}(dq) \\
    & = -\pi i{{\Lambda}^{\mu_1}}_{\nu_1}{{\Lambda}^{\mu_2}}_{\nu_2}\int_{\Gamma}(\eta^{\nu_1\nu_2}-m_W^{-2}q^{\nu_1}q^{\nu_2})\,\Omega_{m_W}(dq) \\
    & = {{\Lambda}^{\mu_1}}_{\nu_1}{{\Lambda}^{\mu_2}}_{\nu_2}D_{m_W}^{\nu_1\nu_2}(\Gamma), 
\end{align*} 
where in line 2 we have used the Lorentz invariance of $\Omega_{m_W}$.
$\Box$

Therefore, since formally
\begin{equation}
D_{m_W}^{\mu_1\mu_2}(\Gamma)=\int_{\Gamma}D_{m_W}^{\mu_1\mu_2}(q)\,dq,
\end{equation}
and
\begin{equation}
D_{m_W}^{\mu_1\mu_2}(\kappa\Gamma)=\int_{\Lambda\Gamma}D_{m_W}^{\mu_1\mu_2}(q)\,dq=\int_{\Gamma}D_{m_W}^{\mu_1\mu_2}(\Lambda q)\,dq,
\end{equation}
we can write
\begin{equation}
D_{m_W}^{\mu_1\mu_2}(\Lambda q)={\Lambda^{\mu_1}}_{\nu_1}{\Lambda^{\mu_2}}_{\nu_2}D_{m_W}^{\nu_1\nu_2}(q).
\end{equation}

\subsection*{Example of muon decay}

Consider muon decay $\mu^{-}\rightarrow e^{-}+\overline{\nu}_e+\nu_{\mu}$, which, by crossing symmetry is equivalent to the process $\mu^{-}+\nu_e\rightarrow e^{-}+\nu_{\mu}$. The Feynman amplitude for this process is
\begin{align*}
{\mathcal M} & = -g_W^2[\overline{u}(e^{-},p_1^{\prime},\alpha_1^{\prime})\gamma_{\mu}(1-\gamma^5)u(\mu^{-},p_1,\alpha_1)] \\
    & \frac{i(-\eta^{\mu\nu}+m_W^{-2}q^{\mu}q^{\nu})}{q^2-m_W^2+i\epsilon} \\
    &  [\overline{u}(\nu_{\mu},p_2^{\prime},\alpha_2^{\prime})\gamma_{\nu}(1-\gamma^5)u(\nu_e,p_2,\alpha_2)], 
\end{align*}
where $q=p_1-p_1^{\prime}$. Thus we write
\begin{equation}
{\mathcal M}_{\alpha_1^{\prime}\alpha_2^{\prime}\alpha_1\alpha_2}(p_1^{\prime},p_2^{\prime},p_1,p_2)=g_W^2{\mathcal M}_{1\mu_1\alpha_1^{\prime}\alpha_1}(p_1^{\prime},p_1)iD_{m_W}^{\mu_1\mu_2}(q){\mathcal M}_{2\mu_2\alpha_2^{\prime}{\alpha_2}}(p_2^{\prime},p_2), 
\end{equation}
where $g_W$ is the weak force coupling constant,
\begin{equation}
{\mathcal M}_{1\mu_1\alpha_1^{\prime}\alpha_1}(p_1^{\prime},p_1)=\overline{u}(e^{-},p_1^{\prime},\alpha_1^{\prime})\gamma_{\mu_1}(1-\gamma^5)u(\mu^{-},p_1,\alpha_1),
\end{equation}
\begin{equation}
{\mathcal M}_{2\mu_2\alpha_2^{\prime}\alpha_2}(p_2^{\prime},p_2)=\overline{u}(\nu_{\mu},p_2^{\prime},\alpha_2^{\prime})\gamma_{\mu_2}(1-\gamma^5)u(\nu_e,p_2,\alpha_2),
\end{equation}
\begin{align*}
u(e^{-},p,\alpha) & = ({\slas p}+m_{e^{-}})e_{\alpha}, \\
u(\mu^{-},p,\alpha) & = ({\slas p}+m_{\mu^{-}})e_{\alpha}, \\
u(\nu_{\mu},p,\alpha) & = ({\slas p}+m_{\nu_{\mu}})e_{\alpha}, \\
u(\nu_e,p,\alpha) & = ({\slas p}+m_{\nu_e})e_{\alpha}, 
\end{align*}
for all $p\in{\bf R}^4,\alpha\in\{0,1,2,3\}$, $e_{\alpha}$ is the $\alpha$th basis element of ${\bf C}^4$ in the standard basis, $m_{e^{-}}$ is the electron mass, $m_{\mu^{-}}$ is the muon mass, $m_{\nu_{\mu}}$ is the mu neutrino mass and $m_{\nu_e}$ is the electron neutrino mass, $q=p_1-p_1^{\prime}=p_2^{\prime}-p_2$ is the momentum transfer and $D_{m_W,\mu_1\mu_2}$ is the W boson propagator (one may also include factors of $m^{-1}$ in the spinors $u(\mbox{type},p,\alpha)$ for normalization).
\newtheorem{lemma}{Lemma}\label{Lemma1}
\begin{lemma}
${{\mathcal M}_{1\mu}^{\alpha^{\prime}}}_{\alpha}$ is a covariant kernel.
\end{lemma}
{\bf Proof}\\
\begin{align*}
{\mathcal M}_{1\mu\alpha^{\prime}\alpha}(p^{\prime}, p) & = \overline{({\slas p^{\prime}}+m_{e^{-}})e_{\alpha^{\prime}}}\gamma_{\mu}(1-\gamma^5)({\slas p}+m_{\mu^{-}})e_{\alpha} \\
    & = e_{\alpha^{\prime}}^{\dagger}(\gamma^0{\slas p^{\prime}}\gamma^0+m_{e^{-}})\gamma^0\gamma_{\mu}(1-\gamma^5)({\slas p}+m_{\mu^{-}})e_{\alpha} \\   
     & = e_{\alpha^{\prime}}^{\dagger}\gamma^0({\slas p^{\prime}}+m_{e^{-}})\gamma_{\mu}(1-\gamma^5)({\slas p}+m_{\mu^{-}})e_{\alpha}.
\end{align*}
Raising the $\alpha^{\prime}$ index using the form $g=\gamma^0$ we obtain
\begin{equation}
{{{\mathcal M}_1}^{\alpha^{\prime}}}_{\mu\alpha}(p^{\prime},p)=e_{\alpha^{\prime}}^{\dagger}({\slas p^{\prime}}+m_{e^{-}})\gamma_{\mu}(1-\gamma^5)({\slas p}+m_{\mu^{-}})e_{\alpha}.
\end{equation}
Therefore
\begin{align*}
{{{\mathcal M}_1}^{\alpha^{\prime}}}_{\mu\alpha}(\kappa p^{\prime},\kappa p) & = e_{\alpha^{\prime}}^{\dagger}(\kappa{\slas p^{\prime}}\kappa^{-1}+m_{e^{-}})\gamma_{\mu}(1-\gamma^5)(\kappa{\slas p}\kappa^{-1}+m_{\mu^{-}})e_{\alpha} \\
    & = e_{\alpha^{\prime}}^{\dagger}\kappa({\slas p^{\prime}}+m_{e^{-}})\kappa^{-1}\gamma_{\mu}(1-\gamma^5)\kappa({\slas p}+m_{\mu^{-}})\kappa^{-1}e_{\alpha} \\
    & = e_{\alpha^{\prime}}^{\dagger}\kappa({\slas p^{\prime}}+m_{e^{-}})\kappa^{-1}\gamma_{\mu}\kappa(1-\gamma^5)({\slas p}+m_{\mu^{-}})\kappa^{-1}e_{\alpha} \\
    & = e_{\alpha^{\prime}}^{\dagger}\kappa({\slas p^{\prime}}+m_{e^{-}}){\Lambda^{-1\nu}}_{\mu}\gamma_{\nu}(1-\gamma^5)({\slas p}+m_{\mu^{-}})\kappa^{-1}e_{\alpha} \\
    & = {\Lambda^{-1\nu}}_{\mu}{\kappa^{\alpha^{\prime}}}_{\beta^{\prime}}{(\kappa^{-1})^{\beta}}_{\alpha}e_{\beta^{\prime}}^{\dagger}\kappa({\slas p^{\prime}}+m_{e^{-}})\gamma_{\nu}(1-\gamma^5)({\slas p}+m_{\mu^{-}})e_{\beta} \\
    & = {\Lambda^{-1\nu}}_{\mu}{\kappa^{\alpha^{\prime}}}_{\beta^{\prime}}{(\kappa^{-1})^{\beta}}_{\alpha}{{{\mathcal M}_1}^{\beta^{\prime}}}_{\nu\beta}(p^{\prime},p), 
\end{align*}
where we have used the fact that any element $\kappa=\left(\begin{array}{cc}
a & 0 \\
0 & a^{\dagger-1}
\end{array}\right)\in K$ commutes with $(1-\gamma^5)=\left(\begin{array}{cc}
0 & 0 \\
0 & 2
\end{array}\right)$ (in the chiral representation) and also the fundamental $K$ intertwining property of the Dirac gamma matrices.
$\Box$

Similarly ${{\mathcal M}_2^{\alpha^{\prime}}}_{\mu\alpha}$ is a covariant kernel. Therefore we can prove the following
\begin{theorem} The Feynman amplitude for muon decay is $K$ covariant.
\end{theorem}
{\bf Proof}\\
We have using Lemma~\ref{Lemma1} and the covariance of $D_{m_W}^{\mu_1\mu_2}$
\begin{align*}
    &  {{\mathcal M}^{\alpha_1^{\prime}\alpha_2^{\prime}}}_{\alpha_1\alpha_2}(\kappa p_1^{\prime},\kappa p_2^{\prime},\kappa p_1,\kappa p_2) = \\
    &  g_{m_W}^2{{\mathcal M}_1^{\alpha_1^{\prime}}}_{\mu_1\alpha_1}(\Lambda p_1^{\prime},\Lambda p_1)iD_{{m_W}}^{\mu_1\mu_2}(\Lambda q){{\mathcal M}_2^{\alpha_2^{\prime}}}_{\mu_2\alpha_2}(\Lambda p_2^{\prime},\Lambda p_2) \\ 
    &  =g_{m_W}^2{\Lambda^{-1\nu_1}}_{\mu_1}{\kappa^{\alpha_1^{\prime}}}_{\beta_1^{\prime}}{{\kappa^{-1}}^{\beta_1}}_{\alpha_1}{{\mathcal M}_1^{\beta_1^{\prime}}}_{\nu_1\beta_1}(p_1^{\prime},p_1){\Lambda^{\mu_1}}_{\sigma_1}{\Lambda^{\mu_2}}_{\sigma_2} \\
    &  iD_{m_W}^{\sigma_1\sigma_2}(q){\Lambda^{-1\nu_2}}_{\mu_2}{\kappa^{\alpha_2^{\prime}}}_{\beta_2^{\prime}}{\kappa^{-1\beta_2}}_{\alpha_2}{{\mathcal M}_2^{\beta_2^{\prime}}}_{\nu_2\beta_2}(p_2^{\prime},p_2) \\
    &  =g_{m_W}^2{\delta^{\nu_1}}_{\sigma_1}{\kappa^{\alpha_1^{\prime}}}_{\beta_1^{\prime}}{{\kappa^{-1}}^{\beta_1}}_{\alpha_1}{{\mathcal M}_1^{\beta_1^{\prime}}}_{\nu_1\beta_1}(p_1^{\prime},p_1) \\
    &  iD_{m_W}^{\sigma_1\sigma_2}(q){\delta^{\nu_2}}_{\sigma_2}{\kappa^{\alpha_2^{\prime}}}_{\beta_2^{\prime}}{\kappa^{-1\beta_2}}_{\alpha_2}{{\mathcal M}_2^{\beta_2^{\prime}}}_{\nu_2\beta_2}(p_2^{\prime},p_2) \\
    &  =g_{m_W}^2{\kappa^{\alpha_1^{\prime}}}_{\beta_1^{\prime}}{{\kappa^{-1}}^{\beta_1}}_{\alpha_1}{{\mathcal M}_1^{\beta_1^{\prime}}}_{\sigma_1\beta_1}(p_1^{\prime},p_1) \\
   &  iD_{m_W}^{\sigma_1\sigma_2}(q){\kappa^{\alpha_2^{\prime}}}_{\beta_2^{\prime}}{\kappa^{-1\beta_2}}_{\alpha_2}{{\mathcal M}_2^{\beta_2^{\prime}}}_{\sigma_2\beta_2}(p_2^{\prime},p_2) \\
   &  ={\kappa^{\alpha_1^{\prime}}}_{\beta_1^{\prime}}{{\kappa^{-1}}^{\beta_1}}_{\alpha_1}{\kappa^{\alpha_2^{\prime}}}_{\beta_2^{\prime}}{\kappa^{-1\beta_2}}_{\alpha_2}{{\mathcal M}^{\beta_1^{\prime}\beta_2^{\prime}}}_{\beta_1\beta_2}(p_1^{\prime},p_2^{\prime},p_1,p_2).  
\end{align*} 
$\Box$

\newtheorem{conjecture}{Conjecture}
\begin{conjecture}
All the Feynman amplitude operators (S matrices) obtained from Feynman diagrams associated with the electroweak force for $(k,l)$ scattering are $K$ intertwining operators from ${\mathcal H}_k$ to ${\mathcal C}_l$.
\end{conjecture}
It is important to note that the condition that an operator is an intertwining operator is a much stronger and more restrictive condition than Lorentz invariance. Conversely we propose that all $K$ intertwining kernels are {\em potentially} associated with real or ``actually occuring" physical processes. It may be possible that the actually ocurring physical processes may be determined by examining the de Rham cohomology of the kernel algebra and its relation to space-time topology. 

\subsection{Higher order computations}

Using similar techniques to those discussed above one can construct many intertwining kernels from covariant measures such as $\Omega_m,\Omega_{f,m}$ for $m\in{\bf R}\setminus\{0\}$, and those associated with fermion loops, the electron self energy  and sums and convolutions  of such covariant measures (it can be shown that such convolutions exists and are $K$ covariant when the test functions are taken from the Schwartz class), and therefore one can compute ``radiative corrections" to M\o ller scattering and compare the results with those obtained using renormalization. One can also construct intertwining operators using the $W$ boson propagator to obtain further understanding of the weak interaction. 

$K$ intertwining kernels ${\mathcal M}\in{\mathcal L}_{kl}$ correspond to potentially occurring scattering processes involving $k$ $|$in$>$ fermions an $l$ $|$out$>$ fermions. It would be of interest to know which kernels correspond to actually occurring scattering processes rather than just potentially occurring processes.

We propose that one can define an analogue of the usual exterior differentiation operation in the kernel algebra bundle, or the superspace bundle with which it is associated, (in which formalism one will define ``creation" and ``annihilation" operators) and then consider an analogue of the de Rham cohomology for this bundle. 
It is to be expected that, as in the finite dimensional case, this cohomology is related to the topology of space-time. This will provide a means for identifying distinguished kernels and hence distinguished, or ``actually occurring" scattering processes. This will be the topic of a subsequent paper.

\section{Conclusion}

We have shown that $(2^{\mbox{nd}}$ quantized) QFT for the electroweak force can be understood by considering the bundle $Q$ as viewed through the representation of its structure group $K$ on the space ${\mathcal H}_1$ of Schwartz spinors, on the Fock space ${\mathcal F}$ of multiparticle states, on the Grassmann algebra ${\mathcal H}$ of physical multiparticle fermionic states and on a number of other algebras. 

Scattering processes with $k$ $|$in$>$ and $l$ $|$out$>$ fermions are described by $K$ intertwining operators (morphisms) $\Xi:{\mathcal H}_k\rightarrow{\mathcal D}_l$ and such operators may be generated by kernels ${\mathcal M}\in{\mathcal L}_{kl}$. ${\mathcal L}=\bigoplus_{k,l=0}^{\infty}{\mathcal L}_{kl}$ forms a kernel algebra and there is a natural mapping from ${\mathcal L}$ to the superalgebra ${\mathcal J}({\mathcal H}_1,{\mathcal D})$.

We propose that it may be possible to describe the strong force in terms of the natural (adjoint) action of $K$ on $su(2,2)$  since $SU(3)$ is closely related to $O(6)$ and $O(6)$ can be obtained from $O(4,2)$ (which is locally isomorphic to $SU(2,2)$) by an analogue of the Wick rotation, and any M\"{o}bius structure can be given a complex structure since the overlap diffeomorphisms are analytic. The strong force will be the subject of a subsequent paper.

\section*{Acknowledgements}
The work described in this paper was supported by Melbourne University and the Commonwealth Scientific and Industrial Research Organisation (CSIRO, Australia). The author is particularly very grateful to Hyam Rubinstein and Sergei Kuzenko for supporting this work and Iain Aitchison for helpful discussions. 

\section*{References}

\rf Arnowitt, R., Chamseddine, A.H. and Nath, P., Supergravity and unification, {\em American Institute of Physics Conference Proceedings} {\bf 116}, 1984, 11-44.

\rf  Barut, A.O. and Doebner, H.-D., {\em Conformal Groups and Related Symmetries}, Springer Verlag, 1986.

\rf Becker, K., Becker, M. and Schwarz, J., {\em String theory and M-theory: A Modern Introduction}, Cambridge University Press, 2007.

\rf Belavin, A. A. and Tarnopolsky, G. M., ``Introduction to string theory and conformal field theory", {\em Physics of Atomic Nuclei}, Vol. 73, No. 5, 2010, pp. 848-877.

\rf Belitsky, A. V. and M\"{u}ller, D., ``Broken conformal invariance and spectrum of anomalous dimensions in QCD", {\em Nuclear Physics B} 537, 1999, 397-442.

\rf Bjorken, J. D. and Drell, S. D., {\em Relativistic Quantum Mechanics}, McGraw-Hill, New York, 1964.

\rf Bjorken, J. D. and Drell, S. D., {\em Relativistic Quantum Fields}, McGraw-Hill, New York, 1965.

\rf Bogolubov, N. N., Logunov, A. A. and Todorov, I. T., {\em Introduction to Axiomatic Quantum Field Theory}, Benjamin, 1975.

\rf  Buchbinder, I. L. and Kuzenko, S. M., {\em Ideas and Methods of Supersymmetry and Supergravity or a Walk Through Superspace}, Institute of Physics Publishing, Bristol and Philadelphia, 1998.

\rf Choquet-Bruhat, Y., DeWitt-Morette, C. and Dillard-Bleick, M., {\em Analysis, Manifolds and Physics}, North-Holland, Amsterdam, 1982.

\rf  Dobrev, V. K., Mack, G., Petkova, V. B., Petrova, S. G. and Todorov, I. T., {\em Harmonic Analysis on the $n$-Dimensional Lorentz Group and its Application to Conformal Quantum Field Theory}, Springer Verlag, 1977.

\rf  Drechsler, W. and Mayer, M. E., {\em Fibre Bundle Techniques in Gauge Theories}, Springer-Verlag, 1977.

\rf Fradkin, E. S. and Palchik, M. Ya., ``Conformal invariance in quantum Yang-Mills theory", {\em Physics Letters} 147 B(1, 2, 3), 1984, 86-90.

\rf  Friedlander, F. G., {\em Introduction to the Theory of Distributions}, Cambridge University Press, 1982.

\rf Gaberdiel, M. R., ``An introduction to conformal field theory", {\em Reports on Progress in Physics} 63, 2000, 607-667.

\rf Green, M. B., Schwarz, J. H. and Witten, E., {em Superstring Theory}, Cambridge University Press, 2012.

\rf Greub, W., Halperin, S. and Vanstone, R., {\em Connections, Curvature and Cohomology}, Vol. I, New York, 1972.

\rf Halmos, P. R., {\em Measure Theory}, Springer-Verlag, New York, 1974.

\rf Itzykson, C. and Zuber, J.-B., {\em Quantum Field Theory}, McGraw-Hill, New York, 1980.

\rf Kirillov, A. A., {\em Elements of the Theory of Representations}, Springer-Verlag, Berlin, 1976.

\rf Kobayashi, S. and Nomizu, K., {\em Foundations of Differential Geometry}, Volume I, Wiley, New York, 1963.

\rf Kol\'{a}r, I., Michor, P. W. and Slov\'{a}k, J., {\em Natural Operations in Differential Geometry}, Springer-Verlag, Berlin, 1993.

\rf Lambert, N., ``M-theory and maximally supersymmetric gauge theories", {\em Annual Review of Nuclear and Particle Science} 62, 2012, 285-313.

\rf Mandl, F. and Shaw, G., {\em Quantum Field Theory}, Wiley, Chichester, 1991.

\rf Mashford, J. S., ``A non-manifold theory of space-time", {\em Journal of Mathematical Physics} 22(9), 1981, 1990-1993.

\rf Mashford, J. S., {\em Invariant measures and M\"{o}bius structures: A framework for field theory}, PhD thesis, University of Melbourne, 2005.

\rf Mashford, J. S., ``An approach to classical quantum field theory based on the geometry of locally conformally flat space-time", 
{\em Advances in Mathematical Physics} (2017), Article ID 8070462, 15 pages, https://doi.org/10.1155/2017/8070462, 2017. 

\rf Meissner, K. A. and Nicolai, H., ``Conformal symmetry and the standard model", {\em Physics Letters B} 648, 2007, 312-317.

\rf Quigg, C., ``Gauge theories of the strong, weak and electromagnetic interactions", Benjamin Cummings, Menlo Park, California, 1983.

\rf Revuz, D., {\em Markov Chains}, North-Holland, Amsterdam, 1984. 

\rf  Rumer, Yu. B., The hydrogen atom and the conformal group, {\em Soviet Physics Doklady} 15(3), 1970.

\rf  Steenrod, N., {\em The Topology of Fibre Bundles}, Princeton University Press, 1951.

\rf Streater, R. F. and Wightman, A. S., {\em PCT, Spin and Statistics, and All That}, Addison-Wesley, 1989.  

\rf Warner, F. W., {\em Foundations of Differentiable Manifolds and Lie Groups}, Springer Verlag, New York, 1983.

\rf  Yosida, K., {\em Functional Analysis}, Springer Verlag, 1980.

\end{document}